\documentclass[aps,prc,twocolumn,showpacs,preprintnumbers,
                          nofootinbib,float,longbibliography]{revtex4-1}
\pdfoutput=1
\usepackage{graphicx, fancybox}
\usepackage{amsmath,amssymb}
\usepackage[usenames, dvipsnames]{color}
\usepackage{soul}
\usepackage{url}
\usepackage{adjustbox}
\usepackage{slashed}
\usepackage{bm}
\usepackage{relsize}
\usepackage{color}
\usepackage{hyperref}
\usepackage{caption}
\usepackage{subcaption}

\newcommand{\beq}{\begin{equation}}
\newcommand{\eeq}{\end{equation}}
\newcommand{\ret}{\mathrm{ret}}
\newcommand{\adv}{\mathrm{adv}}

\newcommand{\sla}{\slashed}

\begin{document}

%\title{Hard probes in an unstable plasma}
\title{Probes of the quark-gluon plasma and plasma instabilities}
\begin{abstract}
Penetrating probes in heavy-ion collisions, like jets and photons, are sensitive to the transport coefficients of the produced quark-gluon plasma, such as shear and bulk viscosity. Quantifying this sensitivity requires a detailed understanding of photon emission and jet-medium interaction in a non-equilibrium plasma. Up to now, such an understanding has been hindered by plasma instabilities which arise out of equilibrium and lead to spurious divergences when evaluating the rate of interaction of hard probes with the plasma. In this paper, we show that taking into account the time evolution of an unstable plasma cures these divergences. We calculate the time evolution of gluon two-point correlators in a setup with small initial momentum anisotropy and show that the gluon occupation density grows exponentially at early times. Based on this calculation, we argue for a phenomenological prescription where instability poles are subtracted. Finally, we show that in the Abelian case instability fields do not affect medium-induced photon emission to our order of approximation. 
\end{abstract}

\author{Sigtryggur Hauksson, Sangyong Jeon, Charles Gale}
\affiliation{Department of Physics, McGill University, 3600 University Street, Montr\'eal, QC, Canada H3A 2T8 \\
{\rm Correspondence}: sigtryggur.hauksson@mail.mcgill.ca}
%\date{\vspace{-5ex}}

\maketitle

\section{Introduction}

In heavy-ion collisions at ultra-relativistic energies a dense medium of quarks and gluons is formed: the quark-gluon plasma (QGP) \cite{Jacak:2012dx}. The medium expands and cools until the quarks and gluons coalesce into soft hadrons which rescatter and fly to the detectors. Remarkably, this time evolution of the QGP is captured by hydrodynamics making the QGP a relativistic fluid \cite{Gale:2013da}. A major goal of heavy-ion collision experiments is to characterize this QGP using transport coefficients, such as shear and bulk viscosity, which quantify its response to weak perturbations and are fundamental properties of QCD.  Hydrodynamic studies have shown that the ratio of shear viscosity to entropy density of QGP is the lowest of any known material \cite{Gale:2013da}, but arguably the precise value is only known within a factor of two or so. Another major goal of these experiments is to understand how the QGP is formed in the first place. Specifically, it needs to be understood how an initial collision of two heavy nuclei at high energies  gives rise to a macroscopic fluid within a time frame of a fm/c or even less. 

Explaining equilibration and transport coefficients in heavy-ion collisions relies on knowledge of the non-equilibrium physics of the quark-gluon plasma. Up until now transport coefficients of the QGP have mostly been extracted by fitting hydrodynamic studies to experimental results of the yield and angular distribution of soft hadrons \cite{Gale:2013da}. An alternative way is offered by hard probes of the QGP, such as photons, jets and heavy quarks. As an example, jets broaden and lose energy as they interact with the QGP medium \cite{Baier:1996sk}. The rate of interaction and its dependence on the energy of a jet particle depends in detail on the makeup of the fluid. Thus the interaction with a thermally equilibrated fluid and a fluid with shear flow will be different, meaning that the energy loss of jets is sensitive to the QGP's shear viscosity \cite{Majumder:2007zh}. To use jets or photons to get the QGP's shear viscosity requires thorough understanding of hard probes in non-equilibrium QGP.

Many issues arise when calculating interaction of hard probes with a non-equilibrium plasma. An important challenge comes from instabilities intrinsic to weakly coupled plasmas. These Weibel instabilities \cite{Moore:2005rp,Mrowczynski:2016etf} come about when quasiparticles that are anisotropically distributed in momentum space radiate soft gluons, the density of which grows exponentially with time. The system is thus intrinsically time dependent. Microscopic calculations of hard probes in non-equilibrium plasma have so far not taken this time dependence into account leading to spurious divergences, such as in the rate of jet particles splitting while interacting with the QGP \cite{Baier:2008js,Romatschke:2006bb,Nopoush:2017zbu,Arnold2002}.

%In this paper we elucidate how one must go beyond the assumption of a static system when evaluating hard probes in a non-equilibrium plasma, see Sec. \ref{sec:correlators}. Using general principles of non-equilibrium quantum field theory we evaluate analytically the time evolution of instability fields at early times when the initial condition is given by a momentum distribution of quasiparticles. We show that including time dependence gives well-behaved rates for jet-medium interaction and photon emission. Thus spurious infinities disappear when the time-dependence of the system is taken into account.    
%
%In Secs. \ref{sec:photon} and \ref{sec:with_medium}, we furthermore study how long-wavelength background fields affect jet-medium interaction and photon emission from the QGP. These background fields can come from instabilities but the calculation is general and is applicable to any initial stage long wavelength fields in heavy-ion collisions. For simplicity we shall only consider the case of Abelian background fields in the paper, leaving the non-Abelian case for future work. 
 
This paper is organized as follows: In Sec. \ref{sec:background} we explain how instabilities in weakly coupled QGP lead to spurious divergences when studying hard probes in a non-equilibrium plasma. In Secs. \ref{sec:main_results} and \ref{sec:correlators} we calculate the time evolution of gluon correlators in a non-equilibrium plasma with slight initial momentum anisotropy. We argue that for phenomenological applications the contribution of instabilities should be subtracted. Finally, in Secs. \ref{sec:photon} and \ref{sec:with_medium} we show that in the Abelian case instability fields do not affect medium-induced photon emission to our order of approximation. Details of calculations are relocated to appendices.
 
\section{Background}
\label{sec:background}

Instabilities in a weakly coupled non-equilibrium quark-gluon plasma lead to spurious divergences when calculating e.g. the rate of photon emission from the plasma or the rate of jet-medium interaction. Understanding the origin of these divergences requires some background on weakly coupled plasmas and quantum-field theoretical calculations of photon emission. 

The ultimate goal when calculating photon emission in a non-equilibrium plasma, is to learn about the QGP formed in experiments by using photons.
This necessitates a flexible approach where rates of photon production can be combined with hydrodynamic simulations of heayv-ion collisions. Specifically, we have two conditions: 
\begin{enumerate}
\item The rate of photon production should only depend on the  properties of the medium in that instant, and not on the medium's history. This requires
\beq
\label{Eq:time_scales_hydro}
t_{\mathrm{process}}\ll t_{\mathrm{medium}} %rename
\eeq
where \(t_{\mathrm{process}}\) is the time it takes to emit a photon and \( t_{\mathrm{medium}} \) is the time scale over which the medium changes substantially. 
\item The rate of photon production should depend solely on macroscopic variables, like pressure and shear flow, that can be obtained from hydrodynamic calculations. This is achieved by describing the medium by a quasiparticle momentum distribution \(f(\mathbf{p})\) that corresponds to the macroscopic variables\footnote{In general, there might be multiple momentum distributions for the same macroscopic variables but the hope is that the calculation is not sensitive to which distribution is chosen as long as the macroscopic variables remain the same.}.
\end{enumerate}

These two conditions have immediate consequences for quantum field theory calculations of photon production. The first condition says that the medium is effectively static during the emission of a photon. We thus want to specify a quasiparticle distribution \(f(\mathbf{p})\) at an initial time \(t_0 \rightarrow - \infty\) which will appear in bare propagators. Assuming that \(f(\mathbf{p})\) remains the same during photon emission, we can use the same bare propagators at all times. Since time ranges from \(-\infty\) to \(\infty\) we can do Fourier transforms and work in frequency space which provides huge simplification. Naively, we expect the results for the rate of photon production to have the same form as in thermal equilibrium, with equilibrium distributions replaced by a more general distribution \(f(\mathbf{p})\)\footnote{A detailed argument is needed to show this \cite{Hauksson2017}, as the original calculation of leading-order photon production in a plasma assumed the Kubo-Martin-Schwinger condition which is only valid in thermal equilibrium \cite{AMYphoton}. }.

Unfortunately, this simple picture does not work in general. As explained in greater detail below, one generally gets a non-sensical, infinite rate of photon production when assuming a static medium characterized by a momentum distribution \(f(\mathbf{p})\).  The culprit are instabilities in the plasma which give rise to rapid exponential growth in the density of soft gluons, violating the assumption of a static medium.  These instabilities arise for any momentum distribution that is anisotropic, i.e.  \(f(\mathbf{p}) \neq f(p)\). (In the case of thermal equilibrium or other isotropic distributions the instabilities are not present and one can assume a static medium.)  In fact, the same problem of divergent rates is present when calculating e.g. the rate of jet-medium interaction \cite{Baier:2008js,Romatschke:2006bb}, heavy-quark potential \cite{Nopoush:2017zbu} and even the rate of interaction among the quasiparticles comprising the medium \cite{Arnold2002}.  

Understanding this problem better requires a detailed discussion of weakly coupled QCD plasmas that are sufficiently close to equilibrium.
Such plasmas are characterized by two energy scales. Firstly, there are quasiparticles -- quarks and gluons -- which are localized and propagate freely, apart from occasionally interacting with each other. Their phase space behaviour can be described by kinetic theory \cite{Arnold2002}, and their distribution functions obey a Boltzmann equation 
\beq
\label{Eq:Boltzmann}
v^{\mu} \frac{\partial f}{\partial x^{\mu}} + \mathbf{F} \cdot \frac{\partial f}{\partial \mathbf{p}} = \overline{C}[f,A]
\eeq
where the  distribution \(f(t,\mathbf{x}; \mathbf{p})\) changes because of external forces \(\mathbf{F}\) and collisions between quasiparticles, as described by \(\overline{C}\). Here colour indices have been suppressed for simplicity.

The quasiparticles radiate gluon fields with energy \(g\Lambda\) where \(g \ll 1\) is the coupling constant. These long-wavelength, soft gluons have high occupancy and can thus be described using classical field theory. Specifically, they obey the classical equations of motion for a gluon field \(A^{\mu}\),
\beq
\label{Eq:class_YM}
\mathcal{D}_{\mu} F^{\mu\nu} = j^{\nu},
\eeq
where \(\mathcal{D}_{\mu}\) is a covariant derivative, \(F^{\mu\nu}\) is the chromoelectromagnetic tensor, and \(j^{\mu}\) is a current which comes from the quark and gluon quasiparticles.

These two coupled equations, Eqs. \eqref{Eq:Boltzmann} and \eqref{Eq:class_YM}, tell us that quasiparticles source gluon fields which deflect the quasiparticles in turn. They can be solved simultaneously, giving rise to an effective field theory for the long-wavelength gluons called Hard Thermal Loops (HTL) \cite{Blaizot2001}.
We write the quasiparticle momentum distribution as
\beq
f(\mathbf{p}) = f_0(\mathbf{p}) + \delta f(x^{\mu};\mathbf{p})
\eeq
where \(\delta f\) is a small fluctuation around the distribution \(f_0\) specified at the initial time \(t_0 \rightarrow - \infty\).
Dropping the subleading collision kernel, Eq. \eqref{Eq:Boltzmann} then becomes
\beq
v^{\mu} \frac{\partial \delta f}{\partial x^{\mu}} = - \mathbf{F} \cdot \frac{\partial f_0}{\partial \mathbf{p}} 
\eeq
where an external force \(\mathbf{F}[A^{\mu}]\) due to an applied gauge field \(A^{\mu}\) sources fluctuation \(\delta f\). Solving for the fluctuation gives a current \(j^{\mu}[A^{\mu}] \sim \int d^3 p \frac{p^{\mu}}{p} \delta f\) which linear response theory tells us is related to the applied field \(A^{\mu}\) through \(j(P) = \Pi_{\ret}(P) A(P)\).
We thus get the retarded self-energy for soft gluons \cite{Mrowczynski2000}
\beq
\label{Eq:ret_self}
\Pi_{\ret}^{\mu\nu}(Q) \sim g^2 \int \frac{d^3p}{2p(2\pi)^3} \;f_0(\mathbf{p}) \left[ g^{\mu\nu} - Q\cdot \partial_{P} \frac{P^{\mu}P^{\nu}}{P\cdot Q-i\epsilon}\right]\Bigg|_{p^0=p}
\eeq 
which depends explicitly on the initial momentum distribution \(f_0\). Here \(P\) and \(Q\) are four-momenta while \(p = \left| \mathbf{p}\right|\) is the three-momentum.
%where again we have omitted color factors and a sum running over the different species in the theory.

Eq. \eqref{Eq:ret_self} contains a wealth of information on how soft gluons propagate in the medium. Continuing with the assumption of a static medium, the retarded propagator \(G^{\mu\nu}_{\ret}(x,y) = \theta(t_x - t_y) \langle [A^{\mu},A^{\nu}]\rangle\) becomes 
\beq
\label{Eq:ret_backg_sec}
G_{\ret}(P) = i \left[P^2 - \Pi_{\ret}(P)\right]^{-1} 
\eeq 
in momentum space where \(\Pi^{\mu\nu}_{\ret}\) is given by Eq. \eqref{Eq:ret_self}. A pole of the retarded propagator, \(\omega = E(\mathbf{p}) - i\Gamma(\mathbf{p})\), contributes 
\beq
\int \frac{d\omega}{2\pi i} \frac{e^{-i\omega t}}{\omega - E + i\Gamma} = e^{-iE t} e^{-\Gamma t} 
\eeq 
in the time domain. This shows that \(\omega = E(\mathbf{p})\) is the dispersion relation of the excitation and \(\Gamma(\mathbf{p})\) is the decay width. 

Whenever the initial momentum distribution \(f_0\) is anistropic, instabilities are present in the system. In \cite{Romatschke2003, Arnold2002} it was shown that a new pole, \(\omega = i\gamma\), appears in the retarded gluon propagator in the upper half complex plane. It corresponds to exponential growth \(e^{\gamma t}\) in soft gluon density in the time domain. This happens as energy is transferred from quasiparticles to the soft chromomagnetic field as it deflect the quasiparticles which source an even stronger field \cite{Mrowczynski:2016etf}. This instability in soft gluon density has been studied extensively numerically, see \cite{Mrowczynski:2016etf} and references therein.

% \begin{figure}
% %    \begin{subfigure}[b]{0.08\textwidth}
%         \begin{center}
%         \begin{minipage}{0.22\textwidth}
%         \centering
%         \includegraphics[width=0.5\textwidth]{./figures/2-to-2_photon/2-to-2}
%         \vspace*{0cm}
%         \end{minipage}
%         \begin{minipage}{0.22\textwidth}
%         \centering
%         \includegraphics[width=0.8\textwidth]{./figures/2-to-2/2-to-2_CG}
%         \vspace*{0cm}
%         \end{minipage}\\
%       \begin{minipage}{0.22\textwidth}
%         \includegraphics[width=\textwidth]{./figures/LPM_photon/LPM_jiggling_CG}
%         \vspace*{0.2cm}
%         \end{minipage}
%       \begin{minipage}{0.22\textwidth}
%       \centering
%         \includegraphics[width=\textwidth]{./figures/LPM_jet/LPM_jiggling_CG}
%         \end{minipage}
%         \vspace*{-3.5cm}
%       \end{center}
%     \caption{Different processes in a weakly coupled quark-gluon plasma: (a) photon production through two-to-two scattering, (b) two-to-two scattering with gluon exchange, (c) photon emission triggered by in-medium interactions, (d) gluon emission triggered by in-medium interactions, . }
% \label{Fig:2to2}
% \end{figure}

\begin{figure}
    \centering
    \begin{subfigure}[b]{0.15\textwidth}
         \centering
         \includegraphics[width=\textwidth]{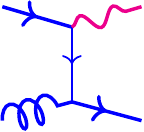}
         \caption{}
         \label{Fig:}
     \end{subfigure}
     \hfill
    \begin{subfigure}[b]{0.15\textwidth}
         \centering
         \includegraphics[width=\textwidth]{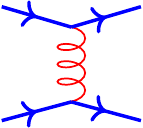}
         \caption{}
         \label{Fig:}
     \end{subfigure} \\
    \begin{subfigure}[b]{0.2\textwidth}
         \centering
         \includegraphics[width=\textwidth]{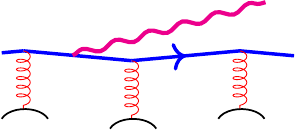}
         \caption{}
         \label{Fig:}
     \end{subfigure}
     \hfill
    \begin{subfigure}[b]{0.2\textwidth}
         \centering
         \includegraphics[width=\textwidth]{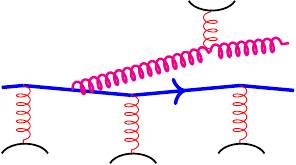}
         \caption{}
         \label{Fig:}
     \end{subfigure}
    \caption{Different processes in a weakly coupled quark-gluon plasma: (a) photon production through two-to-two scattering, (b) two-to-two scattering with gluon exchange, (c) photon emission triggered by in-medium interactions, (d) gluon emission triggered by in-medium interactions}
\label{Fig:2to2}
\end{figure}

The presence of instabilities invalidates the assumption of a static medium. This can for instance be seen when evaluating photon production from the medium. At leading order in the strong coupling constant \(g\), photons are produced through two distinct channels. The first channel is two-to-two scattering with a photon in the final stage, Fig. \ref{Fig:2to2}a, which is unaffected by instabilities in gluon density since the mediator is a quark. \footnote{The interaction of quasiparticles includes two-to-two scattering such as in Fig. \ref{Fig:2to2}b. This channel does not diverge in a static, non-equilibrium medium despite having a gluon mediator. This is because the mediator is a retarded propagator and not an \(rr\) propagator. More physically, the gluon mediator is emitted by the quarks upon interaction and thus does not depend on the accumulated density of gluons in the system.} Its rate has been calculated in a non-equilibrium plasma for various momentum distributions \cite{Baier:1997xc,Schenke:2006yp,Shen:2014nfa}.
%The quark mediator can either have energy \(\sim\Lambda\) or \(\sim g\Lambda\) and in the latter case it is necessary to use its HTL resummed progator. This channel is well-behaved because the quark propagator is insensitive to the high occupancy of gluons due to instabilities. It can thus be calculated assuming a static medium specified by a momentum distribution \(f(\mathbf{p})\). 
The second channel is medium-induced bremsstrahlung of a collinear photon, see Fig. \ref{Fig:2to2}c. A quark is brought slightly off-shell by kicks from the medium's soft gluons, which allows it to emit a photon. The probability for a kick to give the quark transverse momentum \(\mathbf{q}_{\perp}\) is 
\begin{equation}
\label{Eq:coll_kern}
\mathcal{C}(\mathbf{q}_{\perp}) = g^2 C_F \int \frac{dq_0 dq_{z}}{(2\pi)^2} \;2\pi \delta(q_0 - q_{z}) \;  \mathrm{Re}\; G_{rr}(Q)^{\mu\nu} \hat{K}_{\mu} \hat{K}_{\nu}.
\end{equation}
where \(\hat{K}^{\mu} = K^{\mu}/k\) is the direction of the quark \cite{Arnold2002, Hauksson2017}.
Here the crucial ingredient is the correlator \(G^{\mu\nu}_{rr}(x,y) = \frac{1}{2} \langle \left\{ A^{\mu},A^{\nu}]\right\}\) which describes the density of soft excitations. In a static medium with initial time \(t_0 = -\infty\) it is given by 
\beq
G_{rr}(Q) = G_{\ret}\,\Pi_{aa}\, G_{\adv}
\eeq
with \(G_{\adv} = G_{\ret}^{*}\) and \(\Pi_{aa}\) denotes the probability to create the excitation. During emission of a collinear photon, the quark can receive arbitrarily many kicks from the soft gluons. The kicks act coherently and tend to reduce the rate of emission; this is the Landau-Pomeranchuk-Migdal effect \cite{AMYphoton, AMYgluon,Aurenche:2000gf}. Thus in a static medium the rate of photon production through bremsstrahlung has a complicated dependence on \(\mathcal{C}(\mathbf{q}_{\perp})\) which can be seen in Eqs. \eqref{eq:photonrate2} and \eqref{eq:B-like} \cite{Hauksson2017}.

We can now finally see how instabilities invalidate the assumption of a static medium when calculating photon production through medium-induced bremsstrahlung. Roughly speaking the \(rr\) propagator for the instability mode can be approximated as 
\beq
G_{rr} \sim \frac{1}{(q^0 - i\gamma)(q^0 + i\gamma)}.
\eeq
Substituting this into Eq. \eqref{Eq:coll_kern} and ignoring \(q_z\) dependence gives 
\beq
\mathcal{C}(\mathbf{q}_{\perp}) \sim \frac{1}{2\gamma(\mathbf{q}_{\perp})}
\eeq
We thus see that for slowly growing modes \(\gamma \rightarrow 0\) at finite \(\mathbf{q}_{\perp}\) the probability for interacting with soft gluons diverges. This is a sign that our handling of instabilities in a static medium is incorrect.

The function \(\mathcal{C}(\mathbf{q}_{\perp})\) in Eq. \eqref{Eq:coll_kern} not only appears in photon production but also when calculating the rate of jet-medium interaction \cite{AMYgluon,Jeon:2003gi}, as well as interaction of quasiparticles \cite{Arnold2002}. All of these processes thus suffer from the same divergence in a naive calculation in a static medium. Furthermore, a similar problem arises when calculating the imaginary part of the heavy quark potential in a non-equilibrium medium \cite{Nopoush:2017zbu}.

\section{Overview of results and implications for phenomenology}
\label{sec:main_results}

To calculate photon production through bremsstrahlung in a non-equilibrium QGP we must go beyond the assumption of a static medium. Otherwise, we get non-sensical results because of instabilities in soft gluon density. However, including the time evolution of the medium in general is a complicated task, especially since we can no longer do Fourier transforms which are essential to get simple equations for the LPM effect. 
To be able to handle this task, we consider the simplest setup imaginable for the medium and draw lessons from it for more realistic settings. This furthermore gives  a rare opportunity to do analytic calculation in non-equilibrium plasma. 

In our setup the plasma is initially comprised of hard quasiparticles with energy \(\Lambda\)  while soft gluons with energy \(g\Lambda\) are  absent.
% \footnote{They can easily be added as explained in App. \ref{sec:Appendix_KS}}.
The initial condition at \(t_0 = 0\) is given by a slightly anisotropic quasiparticle distribution \(f_0(\mathbf{p})\). 
The anisotropy is defined by
\beq
\xi \sim \frac{\left| \langle p_z\rangle - \langle p_{\perp} \rangle \right|}{\langle p_z\rangle}
\eeq
where \(\langle p_z\rangle, \langle p_{\perp} \rangle\) are the momentum distribution's typical momenta. We assume that \(\xi \ll g\). 
This guarantees that the growth of instabilities is slow enough for us to have handle on the calculation.  We furthermore only consider times shortly after the initial time. This ensures that the density of the soft gluons does not become so high that the HTL approximation is invalidated.

In Sec. \ref{sec:correlators} we calculate the propagators that describe soft gluons in this setup. The retarded correlator becomes
\beq
\label{Eq:ret_final}
G_{\ret}(t_x,t_y; \mathbf{p}) = \int_{\alpha} \frac{dp^0}{2\pi} \;e^{-ip^0(t_x-t_y)} G_{\ret}(p^0,\mathbf{p}).
\eeq
The propagator is written in the time domain where \(t_x, t_y > 0\) are the times of the two fields. Since we assume an infinite spatial extension we can define a three-momentum \(\mathbf{p}\) by Fourier transform. 
The function 
\beq
\label{Eq:ret_main_res}
G_{\ret}(p^0,\mathbf{p}) = \left[(G_{\ret}^0(P))^{-1}- \Pi_{\ret}(P)\right]^{-1}
\eeq 
is the same as in Eq. \eqref{Eq:ret_backg_sec}. It generally 
has poles in the upper half complex plane which correspond to instabilities \cite{Romatschke2003}. Crucially, we must choose a contour \(\alpha\)
 that goes above all poles in the upper half complex plane, as in Fig. \ref{Fig:countour}(a). An instability pole \(p^0 = i\gamma\) then gives \(G_{\ret} \sim e^{\gamma (t_x-t_y)}\) for \(t_x>t_y\) which grows exponentially, showing that the system is unstable to perturbations. Choosing the contour in this way, also guarantees that \(G_{\ret}(t_x,t_y; \mathbf{k}) = 0\) for \(t_y > t_x\).

%
%A non-equilibrium theory exists in the time domain. One can define a frequency space by a Wigner transfrom but that is simply a calculation tool. Our evaluation of Eq. \eqref{Eq:ret_final} is valid at early times when the unstable fields have not grown strong enough. At later times the HTL approximation breaks down. ADD FIG AND SCALES FOR ABELIAN AND NON-ABELIAN. Then we no longer have translational invariance in time and an analytic calculation of correlators is not feasible.
%
%We note that Eq. \eqref{Eq:ret_final} can be viewed as an inverse Laplace transform. This was to be expected since the opposite transform, from the time domain to the frequency domain, can be written like a Laplace transform for a retarded propagator because of the factor \(\theta(t_x-t_y)\).

The important ingredient when calculating photon emission is the \(rr\) correlator of soft gluons which describes the soft gluon density. To find an expression for it we must separate between two scales, namely the soft scale \(g\Lambda\) and the instability growth rate \(\gamma \sim \xi g \Lambda\)\footnote{Strictly speaking, \(\gamma \sim \xi^{3/2}\) but we won't need these more precise estimates in our work \cite{Kurkela:2011ti}}.
As an example we write the retarded correlator in Eq. \eqref{Eq:ret_main_res} as  
\beq
\label{Eq:ret_scale_sep}
G_{\ret}(K) = \widehat{G}_{\ret}(K) + \sum_{i} \frac{A_i}{k^0 - i\gamma_i}
\eeq
where  \(\widehat{G}_{\ret}\) only has poles and branch cuts of order \(g\Lambda\) while \(\gamma_i\) are all poles of order \(\xi g\Lambda\), including instability poles. Using a number of controlled approximations, explained in Sec. \ref{sec:correlators}, we then get the \(rr\) correlator at early times when the gluon occupation density is not so high that the HTL approximation is invalidated. It is  
\beq
\begin{split}
\label{Eq:rr_final}
G_{rr}(t_x,t_y;\mathbf{k}) \approx &\int \frac{dk^0}{2\pi}\;e^{-ik^0(t_x-t_y)} \widehat{G}_{rr}(K) \\ 
&\hspace{-2cm}+ \sum_{i,j} \;\frac{A_i \Pi_{aa}(0) A_j^*}{\gamma_i + \gamma_j} \left[ e^{\gamma_i t_x} e^{\gamma_j t_y}-1\right]
\end{split}
\eeq
with
\beq
\widehat{G}_{rr}(K) = \widehat{G}_{\ret}(K) \,\Pi_{aa}(K)\, \widehat{G}_{\adv}(K) 
\eeq
and \(\widehat{G}_{\adv} = \widehat{G}_{\ret}^{*}\).

The \(rr\) propagator in Eq. \eqref{Eq:rr_final} has a clear physical interpretation. The first term  has no information about the initial time. It is of the same form as the \(rr\) correlator in a static medium, Eq. \eqref{Eq:coll_kern}, except that all instability poles have been subtracted. The second term shows exponential growth due to instabilities at scale \(\gamma \sim \xi g\Lambda\). It vanishes at the initial time \(t_x = t_y = 0\) when the instability modes are not occupied. It is furthermore finite for slow growth rate \(\gamma \rightarrow 0\). 

At very early times the instability part in Eq. \eqref{Eq:rr_final} can be neglected. The probability of a quark to get a transverse kick \(\mathbf{q}_{\perp}\) is then 
\begin{equation}
\label{Eq:coll_kern_correct}
\widehat{\mathcal{C}}(\mathbf{q}_{\perp}) = g^2 C_F \int \frac{dq_0 dq_{z}}{(2\pi)^2} \;2\pi \delta(q_0 - q_{z}) \;  \mathrm{Re}\; \widehat{G}_{rr}(Q)^{\mu\nu} \hat{K}_{\mu} \hat{K}_{\nu}.
\end{equation}
Here all instability poles have been subtracted in \(\widehat{G}_{rr}\) so the probability is finite. As time goes on the occupation density of the soft gluons increases due to instabilities and the second term of Eq. \eqref{Eq:rr_final} must be included. This complicated task is discussed in Sec. \ref{sec:photon}. As even further time passes the HTL approximation used to derive Eq. \eqref{Eq:rr_final} breaks down and numerical calculations are needed to evaluate the evolution of instability modes.

Instability modes have the potential of violating our assumptions for photon production in heavy-ion collisions. The fluctuating  soft gluon cloud in Eq. \eqref{Eq:coll_kern_correct} is sourced  by the hard quasiparticles at each instant so that its effect on photon production only depends on the instantaneous, macroscopic properties of the medium. Conversely, the instability contribution depends on the whole history of the medium and can only be included in phenomenological calculations with great difficulty. 

Fortunately, detailed classical-statistical simulations suggest that plasma instabilities only play a role in the very early stages of heavy-ion collisions \cite{Berges:2013eia,Berges:2013fga}. These calculations describe a weakly coupled, highly occupied classical system with fluctuating initial conditions coming from the color-glass condensate. There the instabilities are important in the approach to a universal, non-equilibrium attractor but once the attractor is reached, detailed information on the initial stages is forgotten and the dynamics is dominated by a turbulent cascade towards higher energies until thermalization is reached. 

 For the phenomenology of photon production in a non-equilibrium QGP, it is therefore reasonable to neglect the contribution of instabilities and use the function \(\widehat{\mathcal{C}}\) in Eq. \eqref{Eq:coll_kern_correct}. This function is non-trivial and has not been calculated fully for a given non-equilibrium distribution. Using it guarantees a finite rate which still includes the essential non-equilibrium information, both from the non-equilibrium quasiparticle distribution \(f\), as well as the soft gluon cloud sourced at each instant by the quasiparticles. The same procedure works for medium-induced jet splitting which also depends on the function \(\widehat{\mathcal{C}}\). We will report on photon production in a non-equilibrium QGP, using this procedure \cite{Hauksson:2018jog, Hauksson2020}.

Applying this prescription, the rate of   emitting photons with momentum \(\mathbf{k}\) through bremsstrahlung is
\beq
\begin{split}
\label{eq:photonrate2}
k \frac{dR}{d^3 k} = \frac{3Q^2 \alpha_{EM}}{4\pi^2} 
 \int \frac{d^3 p}{(2\pi)^3} F(P+K) \left[ 1- F(P)\right] \\
 \times \frac{p^{z\;2} + (p^z+k)^2}{2p^{z\;2} (p^z + k)^2} \;\mathbf{p}_{\perp} \cdot \mathrm{Re}\; \mathbf{f}(\mathbf{p};\mathbf{k})
 \end{split}
\eeq
when the photon is emitted in the \(z\) direction \cite{Hauksson2017}.
Here \(\mathrm{Re}\,\mathbf{f}(\mathbf{p})\) can be thought of as the probability for the quark to gain transverse momentum \(\mathbf{p}\) because of medium kicks. It is solved by the Boltzmann-like equation
\beq 
\label{eq:B-like}
\mathbf{p_{\perp}} = i \delta E\; \mathbf{f}(\mathbf{p_{\perp}}) + \int \frac{d^2 q_{\perp}}{(2\pi)^2}\; \widehat{\mathcal{C}}(\mathbf{q}_{\perp}) \left[\mathbf{f}(\mathbf{p_{\perp}}) - \mathbf{f}(\mathbf{p_{\perp}} + \mathbf{q}_{\perp})\right].
\eeq
where \(\widehat{\mathcal{C}}\) comes from Eq. \eqref{Eq:coll_kern_correct}.

\section{Correlators for unstable fields}
\label{sec:correlators}

We now turn to derive the correlators in Eqs. \eqref{Eq:ret_final} and \eqref{Eq:rr_final} for a weakly anisotropic plasma shortly into its evolution. The retarded propagator is defined by 
\beq
\begin{split}
\label{Eq:ret_equation}
&G_{\ret}(x,y) = G_{\ret}^0(x,y) \\
&+ \int d^{4}z \int d^{4}w \; G_{\ret}^{\,0}(x,z) \Pi_{\ret}(z,w) G_{\ret}(w,y)
\end{split}
\eeq
where \(\Pi_{\ret}\) is the retarded self-energy and \(G_{\ret}^0\) is the bare retarded propagator (see e.g. \cite{Bellac:2011kqa} and Sec. 3 of \cite{Berges:2004yj}). In a static system, such as thermal equilibrium, this equation can be solved by Fourier transforming to the frequency domain, thanks to translational invariance which guarantees that \(G_{\ret}(x,y) = G_{\ret}(x-y)\). We must take a different route to solve Eq. \eqref{Eq:ret_equation} since time translational invariance is broken by instabilities. We assume that our system has infinite spatial extension so that the spatial dependence can be described in Fourier space.

\begin{figure}
\begin{center}
\begin{minipage}{0.15\textwidth}
        \includegraphics[width=0.8\textwidth]{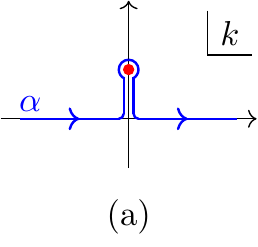}
 \end{minipage}
\begin{minipage}{0.15\textwidth}
\vspace*{-0.16cm}
        \includegraphics[width=0.8\textwidth]{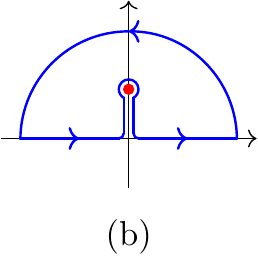}
 \end{minipage}
\begin{minipage}{0.15\textwidth}
        \includegraphics[width=0.8\textwidth]{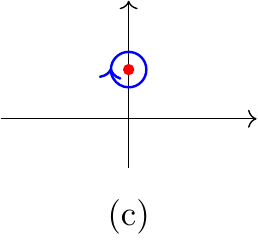}
 \end{minipage}
\end{center}
\caption{Integration contours in the frequency domain. (a) The contour \(\alpha\) runs along the real axis and goes above all poles in the upper half-plane. (b) The countour $\alpha$ continued in the upper half-plane. (c) The contour \(\gamma\) circles all poles in the upper half-plane.}
\label{Fig:countour}
\end{figure}

We start our system at initial time \(t_0 =0\). The time integrals in Eq. \eqref{Eq:ret_equation} range over all times greater than the initial time. Using the properties of retarded functions, we write\footnote{To avoid clutter we denote time coordinates with \(x,y,z,w\) instead of \(x^0,y^0,z^0,w^0\).} 
\beq
\begin{split}
\label{Eq:ret_equation_limits}
&G_{\ret}(x,y) = G_{\ret}^0(x-y) 
\\&+ \int_y^{x} dz \int_y^{z} dw \; G_{\ret}^0(x-z) \,\Pi_{\ret}(z-w) \,G_{\ret}(w,y).
\end{split}
\eeq 
where the dependence on three-momentum is omitted. We have
\(\Pi_{\ret}(z,w)=\Pi_{\ret}(z-w)\) in the HTL approximation, valid at the early times we consider when the soft gluon density is not too high. This equation has the same form as in equilibrium because the initial time does not appear explicitly. Furthermore, \(G_{\ret}(x+\tau,y+\tau)\) is a solution of Eq. \eqref{Eq:ret_equation_limits} for any \(\tau\). This suggests that we can write \(G_{\ret}(x,y) = G_{\ret}(x-y)\). We will therefore try to find a solution\footnote{Here \(k\) can be seen as a frequency coordinate. We will write \(k\) instead of \(k^0\).} 
\beq
\label{Eq:ret_Fourier}
G_{\ret}(x,y) = \int_{\alpha} \frac{dk}{2\pi} \;e^{-ik(x-y)} G_{\ret}(k)
\eeq
for some function \(G_{\ret}(k)\). It's enough to find one such solution because the solution of Eq. \eqref{Eq:ret_equation_limits} is unique. The contour \(\alpha\) goes along the real line and above all instability poles that \(G_{\ret}(k)\) might have in the upper half complex plane, see Fig. \ref{Fig:countour}(a). This ensures that \(G_{\ret}(x,y) = 0\) for \(y>x\).

%\begin{figure}
%   \centering
%    \subcaptionbox{The contour $\alpha$}[0.5\textwidth]{\includegraphics[width=0.15\textwidth]{./figures/alpha_time_contour/alpha_time_contour}}\\
%    \subcaptionbox{The contour $\alpha$ continued in the upper half-plane}[0.5\textwidth]{\includegraphics[width=0.15\textwidth]{./figures/alpha_w_circle/alpha_w_circle}}\\
%    \subcaptionbox{The contour $\gamma$}[0.5\textwidth]{\includegraphics[width=0.15\textwidth]{./figures/gamma_contour/gamma_contour}}
%\caption{Integration contours in the frequency domain. The contour \(\alpha\) runs along the real axis and goes above all poles in the upper half plane. The contour \(\gamma\) circles all poles in the upper half plane. hgf dhasgd hsgfd satyfdusywuyd 8wgx uwy udsyG xuygd usygd usyd }
%\end{figure}

%\begin{figure}
%    \centering
%    \begin{minipage}{0.5\textwidth}
%        \includegraphics[width=0.3\textwidth]{./figures/alpha_time_contour/alpha_time_contour_CG}
%        \caption{The contour \(\alpha\).}
%        \label{Fig:alpha}
%\end{minipage}
%    \quad\quad \\
%    \begin{minipage}{0.5\textwidth}
%        \includegraphics[width=0.3\textwidth]{./figures/alpha_w_circle/alpha_w_circle_CG}
%        \caption{The contour \(\alpha\) continued in the upper half plane.}
%        \label{Fig:alpha_w_circle}
%    \end{minipage} \\
%     \begin{minipage}{0.5\textwidth}
%        \includegraphics[width=0.3\textwidth]{./figures/gamma_contour/gamma_contour_CG}
%        \caption{The contour \(\gamma\).}
%        \label{Fig:gamma}
%    \end{minipage}
%    \caption{Integration contours in the frequency domain. The contour \(\alpha\) runs along the real axis and goes above all poles in the upper half plane. The contour \(\gamma\) circles all poles in the upper half plane.}
%\end{figure}

We will now evaluate the last term in Eq. \eqref{Eq:ret_equation_limits} in detail. Substituting Eq. \eqref{Eq:ret_Fourier} and the Fourier transforms\footnote{Since \(\Pi_{\ret}(x,y) = \Pi_{\ret}(x-y)\), we can define a Fourier transform in the usual way which justifies integrating \(k_2\) over the real line in Eq. \eqref{Eq:ret_subst_freq}. For the full retarded function, a Fourier transform \(G_{\ret}(k) = \int d(x-y)\; e^{ik(x-y)} G_{\ret}(x-y)\) is ill-defined since \(G_{\ret}(x-y)\) has an exponentially growing instability part. Formally, we could define a Laplace transform with the inverse given by Eq. \eqref{Eq:ret_Fourier}. However, we prefer avoiding formal integrals which do not converge. In the end, the time domain is the only physical domain in a non-equilibrium system and \(G_{\ret}(k)\) is just some function that gives the correct retarded function when substituted in Eq. \eqref{Eq:ret_Fourier}. } of \(G_{\ret}^0\) and \(\Pi_{\ret}\) we write that term as
\beq
\begin{split}
\label{Eq:ret_subst_freq}
&\int_y^x dz \int_y^z dw \int \frac{dk_1}{2\pi} \int \frac{dk_2}{2\pi} \int_{\alpha} \frac{dk_3}{2\pi} \;  \\
&e^{-ik_1 (x-z)} e^{-ik_2 (z-w)}e^{-ik_3 (w-y)}\,G_{\ret}^0(k_1)\, \Pi_{\ret}(k_2)\, G_{\ret}(k_3).
\end{split}
\eeq
The time integrals can be done explicitly. This would not be possible if the time integrals were written for all \(z,w\geq0\) since the integral with \(e^{-ik_3 (w-y)}\) would not converge with \(k_3\) in the upper half complex plane. In the end we get 
%\beq
%\label{Eq:ret_eval}
%\begin{split}
%&\int \frac{dk_1}{2\pi} \int \frac{dk_2}{2\pi} \int_{\alpha} \frac{dk_3}{2\pi} \; G_{\ret}^0(k_1)\, \Pi_{\ret}(k_2)\, G_{\ret}(k_3)\\
%&\times\underbrace{\left[-\frac{e^{-ik_1 (x-y)}}{(k_1-k_2)(k_1-k_3)}  -\frac{e^{-ik_2 (x-y)}}{(k_2-k_1)(k_2-k_3)}-\frac{e^{-ik_3 (x-y)}}{(k_3-k_1)(k_3-k_2)}\right]}_{f(k1,k2,k3)} .
%\end{split}
%\eeq
\beq
\label{Eq:ret_eval}
\int \frac{dk_1}{2\pi} \int \frac{dk_2}{2\pi} \int_{\alpha} \frac{dk_3}{2\pi} \; 
G_{\ret}^0(k_1)\, \Pi_{\ret}(k_2)\, G_{\ret}(k_3) f(k_1,k_2,k_3)
\eeq
where
\beq
\begin{split}
f(k_1&,k_2,k_3) = -\frac{e^{-ik_1 (x-y)}}{(k_1-k_2)(k_1-k_3)}  \\
&-\frac{e^{-ik_2 (x-y)}}{(k_2-k_1)(k_2-k_3)}-\frac{e^{-ik_3 (x-y)}}{(k_3-k_1)(k_3-k_2)}
\end{split}
\eeq

Some tricks are needed to evalute Eq. \eqref{Eq:ret_eval}. We notice that the function \(f(k_1,k_2,k_3)\) has no poles in its variables. Thus, we can include a principal value for each term in the function by substituting 
\beq
\begin{split}
\label{Eq:princ_value}
&f(k_1,k_2,k_3) \longrightarrow \\
&\frac{1}{8} \sum_{\substack{\{k_1\rightarrow k_1+i\epsilon_1\}\\\{k_1 \rightarrow k_1-i\epsilon_1\}}} 
\sum_{\substack{\{k_2\rightarrow k_2+i\epsilon_2\}\\\{k_2 \rightarrow k_2-i\epsilon_2\}}}
\sum_{\substack{\{k_3\rightarrow k_3+i\epsilon_3\}\\\{k_3 \rightarrow k_3-i\epsilon_3\}}} f(k_1,k_2,k_3). 
\end{split}
\eeq
Here \(\epsilon_1, \epsilon_2, \epsilon_3>0\) are set to zero in the end. The result must be independent of the order in which they are set to zero. As is shown in App. \ref{sec:Appendix_rr} we can then evaluate the momentum integrals in Eq. \eqref{Eq:ret_eval} using the residue theorem. 
Doing so requires continuing the integration contours to the correct half plane which only contains poles of the function \(f\).
  The final result is 
\beq
\label{Eq:ret_from_appendix}
\begin{split}
\int_{\alpha} &\frac{dk}{2\pi} \;e^{-ik(x-y)} \left[ G_{\ret}(k)  - G^0_{\ret}(k) - G_{\ret}^0(k) \Pi_{\ret}(k) G_{\ret}(k) \right] \\
&= 0.
\end{split}
\eeq 
From this we immediately see that 
\beq
G_{\ret}(k^0,\mathbf{k}) = \left[(G_{\ret}^0(k))^{-1}- \Pi_{\ret}(k)\right]^{-1},
\eeq
confirming our expression for \(G_{\ret}\) in Eq. \eqref{Eq:ret_final}.
We note that the advanced correlator can easily be shown to be 
\beq
\label{Eq:adv_Fourier}
G_{\adv}(x,y) = \int_{\widetilde{\alpha}} \frac{dk}{2\pi} \;e^{-ik(x-y)} G_{\adv}(k).
\eeq  
where \(G_{\adv}(k) = G_{\ret}(k)^*\) and the integration contour is \(\widetilde{\alpha} = \tilde{\alpha}^*\) which goes below all poles of \(G_{\adv}(k)\).

%%%%%%%%%%%%%%%%%%%%%%%%%%%%%%%
%%%%%%%%%%%%%%%%%%%%%%% Continue here!!!

We have found the retarded and advanced correlators. The other  two-point correlator is \(G_{rr} = \frac{1}{2}\langle \{A^{\mu}(x),A^{\nu}(y)\}\rangle\) which gives the occupation density of gluonic modes in the medium. It is
\beq
\label{Eq:rr_integrals}
G_{rr}(x,y) = \int_0^x dz \int_0^y dw\; G_{\ret}(x-z)\, \Pi_{aa}(z-w)\, G_{\adv}(w-y)
\eeq
where the integration limits have been rewritten using properties of the retarded and advanced functions, as well as the initial time \(t_0 = 0\) \cite{Berges:2004yj}.
In general there is an additional term corresponding to correlation with the initial state. Assuming that there are no soft gluons in the initial state, we can omit that term but it could easily be included in our calculations. The integrals depend explicitly on the initial time so we expect that \(G_{rr}(x,y) \neq G_{rr}(x-y)\). Substituing the Fourier transform of the HTL \(\Pi_{aa}\) as well as Eqs. \eqref{Eq:ret_Fourier} and \eqref{Eq:adv_Fourier} gives 
\beq
\label{Eq:rr_prop}
\begin{split}
G&_{rr}(x,y) = \int_{\alpha} \frac{dk_1}{2\pi} \int \frac{dk_2}{2\pi} \int_{\widetilde{\alpha}} \frac{dk_3}{2\pi} \;  \\
&\left[-e^{-ik_2 x}e^{ik_2 y} + e^{-ik_1x}e^{ik_2 y}+e^{-ik_2 x}e^{ik_3 y} - e^{-ik_1 x}e^{ik_3 y}\right] \\
&\times \frac{1}{(k_1-k_2)(k_2-k_3)}\;G_{\ret}(k_1)\, \Pi_{aa}(k_2)\, G_{\adv}(k_3) 
\end{split}
\eeq
after doing the time integrals. 

In order to evaluate the remaining integrals in Eq. \eqref{Eq:rr_prop} we must think about the scales of the problem. The retarded correlator is at two momentum scales: 
\beq
\label{Eq:ret_scale_sep}
G_{\ret}(k) = \widehat{G}_{\ret}(k) + \sum_{i} \frac{A_i}{k - i \gamma_i}
\eeq
Here, \(\widehat{G}_{\ret}\) only has poles and branch cuts of order \(g\Lambda\) which are all in the lower half complex plane while \(\gamma_i\) are all poles of order \(\xi g\Lambda\), with \(\xi \ll g\) the initial anisotropy of the system. We have split \(G_{\ret}\) in a fluctuating part \(\widehat{G}_{\ret}\) that is continually sourced by quasiparticles and an instability part that describes includes time evolution. Similarly, we write
\beq
\label{Eq:adv_scale_sep}
G_{\adv}(k) = \widehat{G}_{\adv}(k) + \sum_{j} \frac{A^*_j}{k + i \gamma_j}
\eeq
where \(\widehat{G}_{\adv} = \widehat{G}_{\ret}^*\). The self-energy \(\Pi_{aa}\) only has poles and branch cuts of order \(g\Lambda\).

\begin{figure}
    \centering
    \hfill
    \begin{subfigure}[b]{0.18\textwidth}
         \centering
         \includegraphics[width=\textwidth]{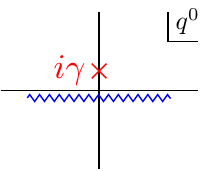}
         \caption{}
         \label{Fig:}
     \end{subfigure}
     \hfill
    \begin{subfigure}[b]{0.18\textwidth}
         \centering
         \includegraphics[width=\textwidth]{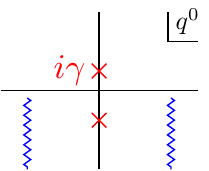}
         \caption{}
         \label{Fig:}
     \end{subfigure}
     \hfill
    \caption{Different branch cuts for Landau damping in retarded gluon correlator: (a) The typical branch cut which spoils a separation of scales, (b) An alternative branch cut which respects separation of scales; if it is chosen additional poles appear in the second Riemann sheet.}
\label{Fig:branch_cuts}
\end{figure}

We need to be careful when writing the retarded correlator as in Eq. \eqref{Eq:ret_scale_sep}. The correlator has a branch cut from \(\omega = -|\mathbf{k}| -i\epsilon\) to \(\omega = |\mathbf{k}| -i\epsilon\) which corresponds to Landau damping. The branch cut is most often chosen to lie just below the real axis but then it will be partially at the scale \(\xi g \Lambda\) which spoils the separation of scales in Eq. \eqref{Eq:ret_scale_sep}.
The remedy is to choose a branch cut that avoids the \(\xi g \Lambda\) region, see Fig. \ref{Fig:branch_cuts}. This results in new decaying modes on the second Riemann sheet \cite{Romatschke2004}. Ultimately, the retarded propagator only exists in the time domain where it is independent of the branch cut we choose.

\begin{figure}
%    \centering
    \includegraphics[width=0.18\textwidth]{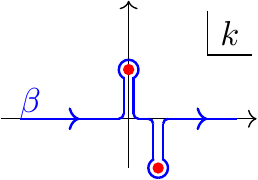}
    \caption{The contour \(\beta\) goes above instability poles of \(G_{\ret}\) in the upper half plane and below instability poles of \(G_{\adv}\) in the lower half plane. Otherwise, it goes along the real axis.}        
        	\label{Fig:beta}
\end{figure}

%%%%%%%%%%%%%%%%%%%%%%%%%%%%%%%%%%%%
We will use controlled approximations to evalute the \(rr\) correlator in Eq. \eqref{Eq:rr_prop}. 
Firstly, we assume that \(x,y \gg 1/g^2 \Lambda\) so that sufficient time has passed since the system was initialized.  
This allows us to  drop any term with \(e^{-i a x}\) where \(\mathrm{Im}\,a <0\) and \(\mathrm{Im}\;a \sim g \Lambda\), as correlations with the initial condition are damped when sufficient time has passed.
 Secondly, we can assume that \(x-y \sim 1/g^2 \Lambda\) since this is the time that medium-induced emission takes. This allows us to drop any term with \(e^{-i a x}\) where \(\mathrm{Re}\,a \sim g\Lambda\), as it oscillates very rapidly during emission and cancels out.
The terms we drop would also be present in a thermally equilibrated system started at an initial time \(t_0=0\). They tell us little about the non-equilibrium physics we are interested in.
%We thus see that poles and branch cuts of the retarded correlator that are \(\sim g\Lambda\) do not contribute when doing the \(k_1\) integral in Eq. \eqref{Eq:rr_mom_int}: They either give  exponential suppression or too rapid oscillations. The same arguments hold for poles and branch cuts \(\sim g\Lambda\) of the advanced correlator.

%%%%%%%%%%%%%%%%%%%%%%%%%%%%%%%%%%

These approximations allow us to to evaluate the \(rr\) propagator at early times. Using the same calculational tricks as before, a lengthy calculation given in App. \ref{sec:Appendix_rr} shows that 
\beq
\label{Eq:rr_result}
\begin{split}
G_{rr}(x,y) \approx &\int \frac{dk}{2\pi}\; \widehat{G}_{\ret}(k) \,\Pi_{aa}(k)\, \widehat{G}_{\adv}(k) \;\;e^{-ik(x-y)} \\
+ \sum_i &\int \frac{dk}{2\pi}\; \frac{A_i}{k-i \gamma_i} \,\Pi_{aa}(k)\, \widehat{G}_{\adv}(k) \;\left(e^{-ikx}-e^{\gamma_i x} \right) e^{iky} \\
+ \sum_j &\int \frac{dk}{2\pi}\; \widehat{G}_{\ret}(k) \,\Pi_{aa}(k)\, \frac{A_j^*}{k+i \gamma_j} \;e^{-ikx} \left( e^{iky}-e^{\gamma_j y}\right) \\
+ \sum_{i,j} &\int \frac{dk}{2\pi}\; \frac{A_i}{k-i \gamma_i} \,\Pi_{aa}(k)\, \frac{A_j^*}{k+i\gamma_j}  \; \\
&\qquad\qquad\times \left(e^{-ikx}-e^{\gamma_i x} \right)\left( e^{iky}-e^{\gamma_j y}\right).
\end{split}
\eeq
where the terms correspond to fluctuating contributions \(k \sim g \Lambda\), instability contributions \(k \sim \xi g \Lambda\) or their cross terms.

Eq. \eqref{Eq:rr_result} has a simple interpretation. Schematically, a mode \(e^{-iEt-\gamma t}\) of the retarded function contributes
\beq
\int^x_{-\infty} dt \;e^{ik^0 t} e^{-iEt-\gamma t} = \frac{-i}{k^0-E+i\gamma} \,e^{i(k^0-E+i\gamma)x}
\eeq 
to the \(rr\) correlator in a system in thermal equilibrium with initial condition at \(t\rightarrow -\infty\). This expression has a pole at \(k^0 = E - i\gamma\).
However, in a non-equilibrium system with initial time at \(t=0\) the corresponding integral is
\beq
\int^x_{0} dt \;e^{ik^0 t} e^{-iEt-\gamma t} =  \frac{-i}{k^0-E+i\gamma} \left[e^{i(k^0-E+i\gamma)x}-1 \right]. 
\eeq
which has no pole. In a similar fashion, there should strictly speaking be no poles in Eq. \eqref{Eq:rr_result}: for a pole \(b \sim g\Lambda\) of \(\widehat{G}_{\ret}\) we should have
\beq
\frac{1}{k-b}\left( e^{-ikx} - e^{-ibx}\right)
\eeq 
Nevertheless, in using our approximations we have dropped all terms \(\sim e^{-ibx}\) since sufficient time has passed to eliminate all traces of an initial time \(t_0=0\). Conversely,
we must retain the analogous factors \(e^{\gamma x}\) for instability modes since they grow exponentially in time.

It is instructive to rewrite Eq. \eqref{Eq:rr_result}. 
We can drop cross-terms between instability and fluctuating modes since the decay or oscillations of fluctuating modes dominates over the slow growth rate of instability terms.\footnote{This can be seen in a simple way. Let's consider a term \(e^{(id+c)t}\) where \(d\sim g\Lambda\) gives oscillations and \(c\sim \xi g \Lambda\) gives  exponential growth. Averaging over the time of interaction in medium-induced emission corresponds to introducing an initial time \(t_0\) which varies over scale \(1/g^2 \Lambda\). This can e.g. be done by integrating \(e^{(id+c)(t-t_0)} e^{-t_0^2/2\sigma^2}\) over the intial time \(t_0\) where the Gaussian with width \(\sigma\sim 1/g^2\Lambda\) corresponds to averaging the time of emission over the time a typical emission takes. Integrating over \(t_0\) then gives a factor \(e^{-\frac{1}{2}\sigma^2(d^2-c^2+2icd)}\) which is heavily suppressed since \(d\gg c\) and \(\sigma d \gg 1\). A full field theoretical calculation gives the same exponential suppression. }
 The last term in Eq. \eqref{Eq:rr_result} has no poles because of the exponentials and can thus be written with a different contour 
\beq
\begin{split}
\label{Eq:rr_instability}
\sum_{i,j} \int_{\beta} \frac{dk}{2\pi} \; &\frac{A_i}{k-i\gamma_i} \,\Pi_{aa}(k)\, \frac{A_j^*}{k+\gamma_j}  \;\\
 &\times\left(e^{-ikx}-e^{\gamma_i x} \right)\left( e^{iky}-e^{\gamma_j y}\right).
\end{split}
\eeq 
Here \(\beta\) is a contour that goes along the real line and above all instability poles of \(G_{\ret}\) in the upper half plane and below all instability poles of \(G_{\adv}\) in the lower half plane, see Fig. \ref{Fig:beta}. Doing the contour integrals then gives
\beq
\begin{split}
\frac{A_i \Pi_{aa}(0) A_j^*}{\gamma_i + \gamma_j} \Big[e^{\gamma_i x} e^{\gamma_j y}&-\theta(x-y) e^{\gamma_i(x-y)}\\
&-\theta(y-x) e^{\gamma_j (y-x)} \Big]
\end{split}
\eeq
where we can ignore poles\footnote{The fact that poles of \(\Pi_{aa}\) can be ignored can be seen as follows: Let's write \(\Pi_{aa}\) as \(A/(k-B)\) where \(B\sim g \Lambda\) is a pole and \(A\) is the residue. Upon performing the contour integral, the pole \(B\) will contribute \(A/(B-a_i)(B-a_j^*) \sim A/(g^2\Lambda)\) while an instability pole will contribute \(A/(a_i-B)(a_i - a_j^*) \sim A/\xi g^2 \Lambda\) which is much bigger. } of \(\Pi_{aa}(k)\) and write \(\Pi_{aa}(a_i) \approx \Pi_{aa}(a_j^*) \approx \Pi_{aa}(0)\). Using that \(a (x-y) \sim \xi/g \ll 1\) gives our final expression for the full \(rr\) correlator which reproduces Eq. \eqref{Eq:rr_final}:
\beq
\begin{split}
G_{rr}(t_x,t_y;\mathbf{k}) \approx &\int \frac{dk^0}{2\pi}\;e^{-ik^0(t_x-t_y)} \\
&\times\widehat{G}_{\ret}(k^0;\mathbf{k}) \,\Pi_{aa}(k^0;\mathbf{k})\, \widehat{G}_{\adv}(k^0;\mathbf{k})  \\ 
&\hspace{-2cm}+ \sum_{i,j} \;\frac{A_i \Pi_{aa}(0) A_j^*}{\gamma_i + \gamma_j} \left[ e^{\gamma_i t_x} e^{\gamma_j t_y}-1\right]
\end{split}
\eeq  
%\beq
%\begin{split}
%\label{Eq:rr_final}
%G_{rr}(t_x,t_y;\mathbf{k}) \approx \int \frac{dk^0}{2\pi}\; \widehat{G}_{\ret}(k^0;\mathbf{k}) \,\Pi_{aa}(k^0)\, \widehat{G}_{\adv}(k) \;\;e^{-ik(x-y)}& \\ 
%+ \sum_{i,j} \;\frac{A_i \Pi_{aa}(0) A_j^*}{\gamma_i + \gamma_j} \left[ e^{\gamma_i x} e^{\gamma_j y}-1\right]&
%\end{split}
%\eeq 
where \(\gamma_i\) and \(A_i\) are functions of the three-momentum \(\mathbf{k}\).

\section{Quark propagators in long-wavelength Abelian background fields}
\label{sec:photon}

\begin{figure}
    \centering
    \includegraphics[width=0.3\textwidth]{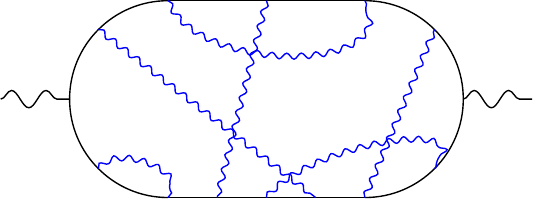}
    \caption{An example of diagrams that are summed up to evaluate photon production in an Abelian background field. The thin lines joining quark propagators are background field insertions.} 
	\label{Fig:crossed_diagr}       
\end{figure}

\begin{figure}
    \centering
    \includegraphics[width=0.3\textwidth]{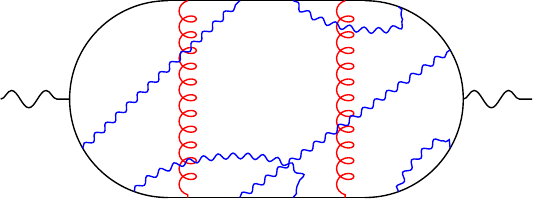}
    \caption{A diagram for medium-induced photon production in the presence of a background field. The red gluons denote medium kicks at energy \(g\Lambda\) which are time ordered. The blue lines denote kicks from the background field at energy \(\xi g \Lambda\). They are not time ordered.}   
	\label{Fig:crossed_HTL}     
\end{figure}

We have argued that for phenomenological applications the time-dependent instability field in Eq. \eqref{Eq:rr_final} should simply be subtracted, leaving a simple expression for photon production in a non-equilibrium plasma. It is nevertheless interesting to explore the effect of the long-wavelength instability field, both from a theoretical point of view, as well as as a first step towards including classical fields at the early stages of heavy-ion collisions.

We will now calculate how the long-wavelength background fields modify photon emission, focusing on the case of an Abelian plasma. In particular we consider how the background fields modify medium-induced bremsstrahlung as seen in Fig. \ref{Fig:2to2}c which suffers from spurious divergences when one assumes a static, non-equilibrium plasma. Our setup is fairly general: The medium can be described by the \(rr\) correlator in Eq. \eqref{Eq:rr_final} but also by any other \(rr\) correlator which has two different scales, fluctuating time-independent excitations with energy \(g\Lambda\) and a time-dependent background field with energy \(\xi g \Lambda\), \(\xi\ll g\). This calculation also extends easily to jet-medium interaction and quasiparticle splitting as seen in Fig. \ref{Fig:2to2}d. Our goal is to sum up non-perturbative effects of the background field at a given order in \(l\Delta t \ll 1\) where \(l \sim \xi g \Lambda\) is the small momentum of the background field and \(\Delta t \sim 1/g^2 \Lambda\) is the time photon emission takes. 

The two energy scales, i.e. the fluctuating field at \(g\Lambda\) and the background field at \(\xi g\Lambda\), affect photon emission in very different ways. The time for collinear bremsstrahlung of photons is \(\sim 1/g^2\Lambda\) which is very long compared to the time \(1/g\Lambda\) for a typical medium kick. Thus the medium-kicks are ordered in time and diagrams with crossed rungs like in Fig. \ref{Fig:crossed_diagr} are suppressed. On the other hand the long-wavelength background field has wavelength \(\sim 1/\xi g \Lambda\) which is much longer than the time for photon emission. Thus we must evaluate diagrams like in Fig. \ref{Fig:crossed_diagr} for the background field. These diagrams are complicated because of the color factors and can only realistically be summed up in the case of an Abelian background field or a non-Abelian background field in the large \(N_c\) limit where only planar diagrams contribute. We focus on the Abelian case here. Our goal is to do a calculation that includes both medium kicks and the background field as can be seen in Fig. \ref{Fig:crossed_HTL}.

We make a few assumptions about the scales of the problem. Firstly, we assume that the momentum \(l\) of the long-wavelength background field satisfies
\beq
\label{Eq:assumptions1}
l \Delta t \ll 1
\eeq
where \(\Delta t\) is the time the emission of a photon takes. In our case \(\Delta t \sim 1/g^2\Lambda\) so for instability fields \(\xi \ll g\). Furthermore, we assume that \(\gamma \Delta t \ll 1\) where \(1/\gamma\) is the time over which the background fields change appreciably.  We also assume 
\beq
\label{Eq:lambda_deltat}
1 \ll \Lambda \Delta t
\eeq
where \(\Lambda\) is the hard scale of the medium and \(\Delta t\) is the time an emission takes. Medium-induced emission of photons or gluons takes time \(\sim 1/g^2 \Lambda\) which is long enough to fulfill the condition. In general Eq. \eqref{Eq:lambda_deltat} is satisfied for off-shell photon emission with virtuality \(Q^2 \ll \Lambda^2\).

We finally assume that the wavelength of the background fields cannot be so long that it correlates two subsequent gluon emissions. In other words
\beq
\label{Eq:assumptions2}
\frac{1}{\Lambda \left(\Delta t\right)^2} \ll l
\eeq
where \(1/\Lambda \left(\Delta t\right)^2 \sim \Lambda/g^4\), the mean free path for gluon emission. 

\begin{figure}
    \centering
    \includegraphics[width=0.2\textwidth]{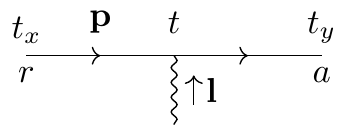}
    \caption{A retarded propagator with one background field insertion.} 
  	\label{Fig:one_background}      
\end{figure}

Quark propagators are modified in the presence of background fields. The bare retarded propagator is 
\beq
S_{\ret}^{(0)}(t_x,t_y;\mathbf{p}) = \frac{1}{2}\theta(\Delta t) \left[ e^{-ip\Delta t} \sla{\widehat{P}} - e^{ip\Delta t} \sla{\widetilde{P}}\right]
\eeq
where \(\Delta t = t_x - t_y\) and  \(\widehat{P} = (1,\widehat{\mathbf{p}})\) and  \(\widetilde{P} = (-1,\widehat{\mathbf{p}})\) denote different polarizations. Adding one background field insertion, Fig. \ref{Fig:one_background}, gives
\beq
\begin{split}
S^{(1)}_{\ret}(t_x,t_y;\mathbf{p}; \mathbf{l}) &= igA_{\mu} \int dt\; S^{(0)}_{\ret}(t_x,t;\mathbf{p}) \gamma^{\mu} S^{(0)}_{\ret}(t,t_y;\mathbf{p} + \mathbf{l})%\\
%&= ig A_{\mu} \frac{1}{4} \theta(t_x -t_y)\int_{t_y}^{t_x} dt \;  \left[e^{-ip(t_x-t)} \slashed{\widehat{P}} - e^{ip(t_x-t)} \slashed{\widetilde{P}} \right] \gamma^{\mu} \left[e^{-i\left|\mathbf{p}+\mathbf{l} \right|(t-t_y)} \slashed{\widehat{P}} - e^{i\left|\mathbf{p}+\mathbf{l} \right|(t-t_y)} \slashed{\widetilde{P}} \right]
\end{split}
\eeq
We can take the background field \(A^{\mu}\) out of the time integral since it changes slowly.
This gives 
%\beq
%\begin{split}
%S^{(1)}_{\ret}(t_x,t_y; \mathbf{p}; \mathbf{l}) \approx \frac{1}{2} \theta(t_x-t_y) e^{ip(t_x-t_y)} \Big[ (ig A_{\mu} \widehat{P}^{\mu})  \sla{\widehat{P}} \frac{e^{-i\,\widehat{\mathbf{p}}\cdot\mathbf{l}(t_x-t_y)}-1}{-i\,\widehat{\mathbf{p}}\cdot\mathbf{l}}& \\
% +\,(ig A_{\mu} \widetilde{P}^{\mu})  \sla{\widetilde{P}}\frac{e^{i\,\widehat{\mathbf{p}}\cdot\mathbf{l}(t_x-t_y)}-1}{i\,\widehat{\mathbf{p}}\cdot\mathbf{l}} &\Big]
%\end{split}
%\eeq
\beq
\label{Eq:ret_one_insertion}
\begin{split}
S^{(1)}_{\ret}(t_x,t_y; \mathbf{p}; \mathbf{l}) \approx \qquad\qquad\qquad\qquad\qquad\qquad&\\
\frac{1}{2} \theta(\Delta t) e^{ip\Delta t} \Big[ (ig A_{\mu} \widehat{P}^{\mu}) \;\sla{\widehat{P}}\;\left( \Delta t -\frac{1}{2} i\,  l_{||}  (\Delta t)^2  \right)& \\
 +\,(ig A_{\mu} \widetilde{P}^{\mu})\;\sla{\widetilde{P}}\; \left(  \Delta t +\frac{1}{2} i\,  l_{||}  (\Delta t)^2 \right) &\Big]
\end{split}
\eeq
Here we have expanded in \(\Delta t \,l_{||} \ll 1\) with \(l_{||} = \widehat{\mathbf{p}}\cdot\mathbf{l}\). 
Terms with \(\Delta t\)  in Eq. \eqref{Eq:ret_one_insertion} denote a potential phase rotation in the background field. The subleading term with \( l_{||}  (\Delta t)^2\), gives the first derivative of the background field \(A^{\mu}\) and thus denotes the effect of electromagnetic fields on photon emission. Higher order terms are not amenable to evaluation using our methods. \footnote{The omitted terms in Eqs. \eqref{Eq:ret_one_insertion} and \eqref{Eq:instab_perm} are in fact subleading. Cross terms like \(\sla{\widehat{P}} \gamma^{\mu} \sla{\widetilde{P}}\) which denote spin flip in the background field give \(\frac{1}{p} \ll l (\Delta t)^2\) after doing the time integral and can thus be ignored. Furthermore, we can ignore spin precession in the background fields. It will give spinor factors with \(\widehat{P\!+\!L}^{\mu} \approx P^{\mu} +\mathcal{O}(l/\Lambda)\). After doing the time integral the spin precession correction gives a term \(\mathcal{O}(l\Delta t/\Lambda)\) which is subleading to the terms in Eq. \eqref{Eq:instab_perm}. }

The retarded quark propagator with an arbitrary number of background field insertions is
\beq
\label{Eq:instab_perm}
\begin{split}
&S^{(n)}_{\ret}(t_x,t_y;\mathbf{p}; \{\mathbf{l}_1,\dots,\mathbf{l}_n\}) \approx \\
& \frac{1}{2} \theta(\Delta t) \Big[ (ig A_{\mu} \widehat{P}^{\mu})^n e^{-ip\Delta t} \sla{\widehat{P}} \prod_{i=1}^n \frac{e^{-i\,l_i^{||}\Delta t}-1}{-i\,l_i^{||}} \\
 &\qquad\quad+\,(-1)^{n+1}(ig A_{\mu} \widetilde{P}^{\mu})^n e^{ip\Delta t} \sla{\widetilde{P}} \prod_{i=1}^n \frac{e^{i\,l_i^{||}\Delta t}-1}{i\,l_i^{||}} \Big]
\end{split}
\eeq
This simple form is achieved by summing over all the different permutations of attaching  \(n\) background field insertions. The analogous expression for the advanced propagator has an overall minus sign and \(\theta(-\Delta t)\) instead of \(\theta(\Delta t)\). 

Eq. \eqref{Eq:instab_perm} can be derived by noting that the dependence on background field momentum is
\beq
\begin{split}
& \int \frac{d\omega}{2\pi} \;e^{-i\omega\Delta t} \;\frac{i}{\omega-p+i\epsilon}\;\frac{i}{\omega-\left| \mathbf{p} \!+ \!\mathbf{l}_1\right|+i\epsilon}\dots \\
 &\qquad\qquad\qquad\times\frac{i}{\omega-\left| \mathbf{p} \!+\! \mathbf{l}_1\!+\!\dots\!+\!\mathbf{l}_n\right|+i\epsilon}.
\end{split}
\eeq
for \(n\) ordered instability insertions. Performing the integral and expanding in \(l_{i\;||} = \hat{\mathbf{p}} \cdot \mathbf{l}_i\) gives a complicated expression. It is hugely simplified by summing over all permutations of attaching  \(n\) background field insertions, and using that 
\beq
%\label{Eq:perm_ident}
\sum_{\substack{\mathrm{permute} \\ \{l_1,\dots,l_j\}}} \frac{1}{(l_1+\dots +l_j)(l_2+\dots +l_j)\dots l_j} = \frac{1}{l_1\dots l_j}
\eeq

\begin{figure}
    \centering
    \includegraphics[width=0.45\textwidth]{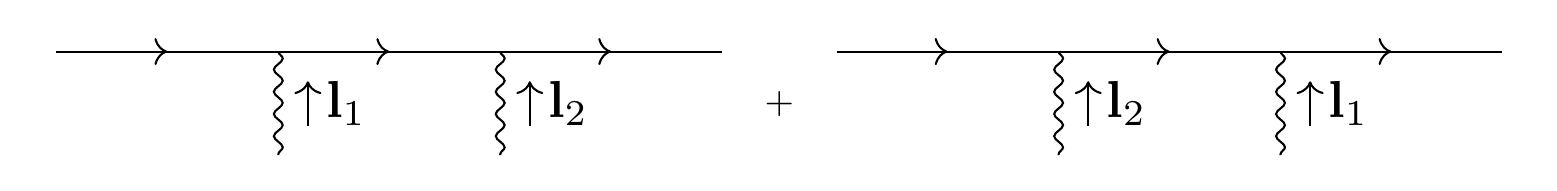}
    \caption{A retarded propagator with two background field insertions. The different ways of attaching the insertions are summed over.} 
    	\label{Fig:two_background}       
\end{figure}

% The complicated expression in Eq. \eqref{Eq:S_n_bef_perm_3} takes a simple form after summing over all possible ways of ordering the different background field insertions, see Fig. \ref{Fig:two_background} for an example. We permute the insertions in two steps. Firstly,  for each term in the sum over \(j\) we do permutations within \(\{b_1,\dots,b_j\}\) and \(b_{j+1},\dots,b_n\) seperately. This is straightforward since
% \beq
% %\label{Eq:perm_ident}
% \sum_{\substack{\mathrm{permute} \\ \{l_1,\dots,l_j\}}} \frac{1}{(l_1+\dots +l_j)(l_2+\dots +l_j)\dots l_j} = \frac{1}{l_1\dots l_j}
% \eeq
% as can be shown by induction. Secondly, we sum over all different ways of splitting \(\{l_1,\dots,l_n\}\) into two sets with \(j\) and \(n-j\) elements. This gives that summing over all permutations in Eq. \eqref{Eq:S_n_bef_perm_3} leads to
% \beq
% \begin{split}
% &S^{(n)}_{\ret}(t_x,t_y;\mathbf{p}; \left\{\mathbf{l}_1,\dots,\mathbf{l}_n\right\})  \sim \\
% &(-i)^n  \theta(\Delta t)\, e^{-ip\Delta t}\;\frac{1}{l_1\dots l_n} 
% \sum_{j=0}^n (-1)^j \sum_{\substack{(b_1,\dots,b_j) \\ \subset (l_1,\dots,l_n)}} e^{-i(b_1+\dots+b_i)(t_x-t_y)} \\
% &\qquad\qquad=  \theta(\Delta t)\, e^{-ip\Delta t} \prod_{j=1}^n \frac{e^{-i l_j\Delta t}-1}{-i l_j}
% \end{split}
% \eeq 

\begin{figure*}[ht]
    \centering
   \includegraphics[width=0.8\textwidth]{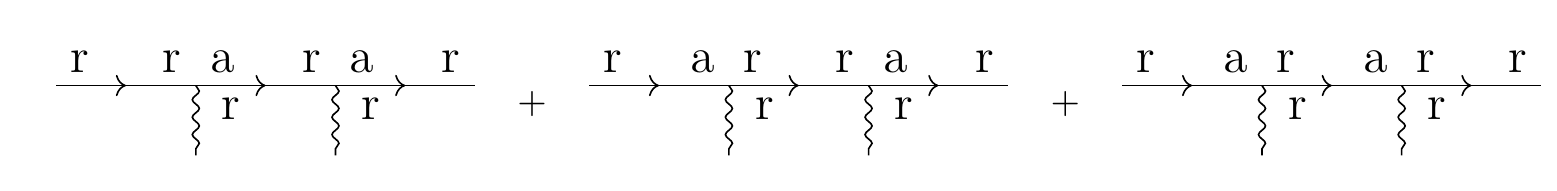}
    \caption{The three different ways of ordering \(r\) and \(a\) indices on a \(rr\) propagator with two background field insertions.}   
    	\label{Fig:rr_indices}   
\end{figure*}

Finally, we must evaluate how the \(rr\) propagator is modified in the presence of a long-wavelength background field. The bare \(rr\) propagator in the time domain is
\beq
\label{Eq:rr_bare}
S^{(0)}_{rr}(t_x,t_y;\mathbf{k}) = \left[\frac{1}{2} - F_q(\mathbf{k}) \right] \left( S^{(0)}_{\ret}(t_x,t_y;\mathbf{k}) - S^{(0)}_{\adv}(t_x,t_y;\mathbf{k})\right)
\eeq 
where
\beq
F(\mathbf{k}) = \left\{ \begin{array}{lr}
f_q(\mathbf{k}), & \text{for } \widehat{K} \\
1-f_{\bar{q}}(-\mathbf{k}), & \text{for } \widetilde{K} \end{array} \right.
\eeq
 describes the momentum distribution for incoming quarks and outgoing antiquarks, respectively. There are many ways to add background field insertions in the \(ra\) basis. As an example Fig. \ref{Fig:rr_indices} shows the three possible ways of including two background field insertions. To find them we have used that a background field insertion has index \(r\), that each vertex has an odd number of \(a\) indices and that bare \(aa\) propagators vanish \cite{AMYphoton}.
%  , see App. \ref{sec:Appendix_KS}.
%  Schematically, these contributions can be written as 
% \beq
% S^{(0)}_{rr}S^{(0)}_{ar} S^{(0)}_{ar} + S^{(0)}_{ra}S^{(0)}_{rr} S^{(0)}_{ar}  + S^{(0)}_{ra}S^{(0)}_{ra} S^{(0)}_{rr}
% \eeq 
% where, say, the first term has momenta
% \beq
% S^{(0)}_{rr}(t_x,t_1;\mathbf{p})S^{(0)}_{ar}(t_1,t_2;\mathbf{p} + \mathbf{l}_1) S^{(0)}_{ar}(t_2,t_y;\mathbf{p} + \mathbf{l}_1 + \mathbf{l}_2).
% \eeq 
% Using Eq. \eqref{Eq:rr_bare}, these three contributions can be written as 
% \beq
% \begin{split}
% &\left[ \frac{1}{2} - F_q(\mathbf{k})\right]\left[S^{(0)}_{ra} - S^{(0)}_{ar}\right]S^{(0)}_{ar} S^{(0)}_{ar} + \\
% &\left[\frac{1}{2} - F_q(\mathbf{k}+\mathbf{l}_1) \right] S^{(0)}_{ra}\left[S^{(0)}_{ra} - S^{(0)}_{ar} \right]S^{(0)}_{ar}  + \\
% &\left[\frac{1}{2} - F_q(\mathbf{k}+\mathbf{l}_1+\mathbf{l}_2) \right] S^{(0)}_{ra} S^{(0)}_{ra} \left[ S^{(0)}_{ra} - S^{(0)}_{ar} \right]
% \end{split}
% \eeq
Assuming that the momentum distributions in each propagator are the same to our order of approximation, \(f_q(\mathbf{k}) \approx f_q(\mathbf{k}+\mathbf{l}_1) \approx f_q(\mathbf{k}+\mathbf{l}_1+\mathbf{l}_2) \)\footnote{By making the approximation \(f_q(\mathbf{k}+ \mathbf{l}) \approx f_q(\mathbf{k})\) we ignore how quarks are rotated in the background field during emission. This correction is of order \(\mathbf{l} \cdot \nabla f(\mathbf{p}) \sim \frac{l}{\Lambda} f(\mathbf{p})\). Such terms have a combination of retarded and advanced propagator with no simple time ordering. The time integral at the vertex with momentum contribution \(\mathbf{l}\) will thus give \(T\), the time that has passed since the initial conditions that specified the momentum distribution \(f(\mathbf{p})\).
% \footnote{This is of course not the same initial time as was used in Sec. \ref{sec:correlators} when deriving Eq. \eqref{Eq:rr_final}. There is nothing stopping us from using a different initial time here.}  
Choosing \(T \gtrsim \Delta t\) so that the momentum distribution describes the quarks just before they emit the photon, it is easy to see that the correction is subleading to Eq. \eqref{Eq:instab_perm}.}, and using Eq. \eqref{Eq:rr_bare}, most of the terms cancel \cite{Hauksson2017}. We end up with 
\beq
S^{(2)}_{rr}= \left[\frac{1}{2} -   F_q(\mathbf{k})\right] \left( S^{(2)}_{\ret}  - S^{(2)}_{\adv} \right).
\eeq
A similar cancellation takes place for any number of background field insertions so that in the end
\beq
\begin{split}
\label{Eq:rr_F}
&S^{(n)}_{rr}(t_x,t_y;\mathbf{p}; \{\mathbf{l}_1,\dots,\mathbf{l}_n\}) = \\
&\left[\frac{1}{2} -   F(\mathbf{k})\right] \Big( S^{(n)}_{\ret}(t_x,t_y;\mathbf{p};\{\mathbf{l}_1,\dots,\mathbf{l}_n\}) \\
&\qquad\qquad\quad - S^{(n)}_{\adv}(t_x,t_y;\mathbf{p}; \{\mathbf{l}_1,\dots,\mathbf{l}_n\}) \Big).
\end{split}
\eeq

%\begin{widetext}
%\onecolumngrid
%\twocolumngrid
%\end{widetext} 

\section{Medium-induced photon emission in Abelian background fields}
\label{sec:with_medium}

\begin{figure}
    \centering
    \includegraphics[width=0.27\textwidth]{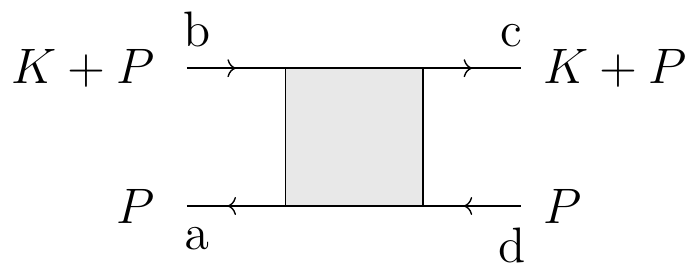}
    \caption{Definition of the four point quark function \(S_{abcd}\).}    
    	\label{Fig:4_pt_fun}    
\end{figure}

We turn to evaluating photon emission in an Abelian background field. For simplicity, we begin by only considering the long wavelength background field with momentum \(\xi g \Lambda\), considering medium kicks with momentum \(g\Lambda\) below. 

\begin{figure}
    \centering
    \includegraphics[width=0.15\textwidth]{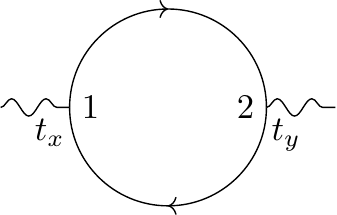}
    \caption{Photon self-energy diagram, without a background field or a medium.}  
	\label{Fig:bare_quark_loop}      
\end{figure}

On-shell photon emission from on-shell quarks is kinematically suppressed in the absence of kicks from a background field or a medium. The rate of on-shell photon emission is given by \(\Pi_{12}^{\gamma}\) which goes like the four-point quark correlator \(S_{1122}\), see Fig. \ref{Fig:4_pt_fun}. We show in Appendix \ref{sec:Appendix_distr} that
\beq
\label{Eq:S_1122}
S_{1122} = 2 F(P+K) \left(1-F(P)\right) \mathrm{Re}\; S_{rraa}
\eeq
where we have gone to the \(ra\) basis in the closed-time path formalism \cite{Bellac:2011kqa} defined by 
\beq
\psi_r = \frac{\psi_1 + \psi_2}{2}; \;\;\; \psi_a = \psi_1 - \psi_2.
\eeq
The momentum factors in Eq. \eqref{Eq:S_1122} describe different channels. As an example with \(p^0 > 0\) we get \(f_q(\mathbf{p}+\mathbf{k}) \left( 1-f_q(\mathbf{p})\right)\) which denotes a quark with momentum \(\mathbf{p} + \mathbf{k}\) emitting a photon with momentum \(\mathbf{k}\) through bremsstrahlung. The rate of emitting an on-shell photon with momentum \(k\) through quark bremsstrahlung then goes like
% We focus on the channel where a quark emits a photon through bremsstrahlung. In the absence of a background field the photon polarization tensor in the time domain is 
\beq
\begin{split}
i &\Pi_{12\;\mu}^{\mu}(\mathbf{k}) = \frac{e^2}{4} \,\mathrm{Tr} \left[ \gamma_{\mu} \sla{\widehat{K}} \gamma^{\mu} \sla{\widehat{K}}\right] 2 f_q(\mathbf{p} + \mathbf{k}) (1-f_q(\mathbf{p})) 
\\&\times\mathrm{Re}  \int d(t_x-t_y) e^{ik(t_x-t_y)} \theta(t_x-t_y) \;e^{-i(\left|\mathbf{p}+\mathbf{k} \right|-p)(t_x-t_y)} 
\end{split}
\eeq
%(CHECK FACTORS!)
as can be seen in Fig. \ref{Fig:bare_quark_loop}. 
% For this channel we choose the part of the retarded propagator with \(\widehat{P\!+\!K} \approx \widehat{K}\) and the advanced propagator with \(\widehat{P} \approx \widehat{K}\) since the emission is collinear.
% The time dependence is by 
% \beq
% \label{Eq:B_0}
% B_0 := \theta(t_x-t_y)e^{-i(\left|\mathbf{p}+\mathbf{k} \right|-p)(t_x-t_y)} 
% \eeq
The frequency integral gives \(\delta(k-\left|\mathbf{p}+\mathbf{k} \right|+p)\)
% \beq
% \mathrm{Re}\; \int d(t_x-t_y) e^{ik^0(t_x-t_y)} B_0 = -2\pi \delta(k^0-\left|\mathbf{p}+\mathbf{k} \right|+p)
% \eeq
which vanishes under integration over \(\mathbf{p}\) for an on-shell photon. This is simply because the emission is kinematically suppressed. 

% The photon polarization tensor we need to evaluate is \(\Pi_{12}^{\gamma}\) which goes like the four-point quark propagator \(S_{1122}\), see Fig. \ref{Fig:4_pt_fun}. To use the propagators we have derived we must go to the \(ra\) basis. We show in Appendix \ref{sec:Appendix_distr} that
% \beq
% \label{Eq:S_1122}
% S_{1122} = 2 F(P+K) \left(1-F(P)\right) \mathrm{Re}\; S_{rraa}
% \eeq
% The momentum factors describe the three different channels. We can firsly have \(f_q(\mathbf{p}+\mathbf{k}) \left( 1-f_q(\mathbf{p})\right)\) when a quark emits a photon through bremsstrahlung. Secondly, \(f_{\bar{q}}(-\mathbf{p}) \left( 1-f_{\bar{q}}(-\mathbf{p}-\mathbf{k})\right)\) describes an antiquark emitting a photon through bremsstrahlung. Finally, \(f_q(\mathbf{p}+\mathbf{k}) f_{\bar{q}}(-\mathbf{p})\) desribes pair annihilation of a quark and an antiquark.   

% We might however expect that the background fields bring the quarks sufficiently off-shell to allow for photon emission. 

We now turn on the background field and see whether on-shell photon emission becomes possible. Since we have assumed that \(\langle A\rangle = 0\) but \(G_{rr} = \frac{1}{2}\langle \{ A,A\} \rangle \neq 0\) we must pair up the background field insertions into \(rr\) two-point functions to account for fluctuations in the background field. An example of a contributing can be seen in Fig. \ref{Fig:crossed_diagr}.

The upper quark rail with momentum \(\mathbf{p}+\mathbf{k}\) becomes 
\beq
\label{Eq:upper_rail}
\begin{split}
S_{\ret}^{n_1}&(t_x,t_y;\mathbf{p}+\mathbf{k}; \left\{\mathbf{l}_1,\dots \right\}) = S_{\ret}^{(0)}(t_x,t_y;\mathbf{p}+\mathbf{k}) \\
&\times(igA_{\mu}\widehat{K}^{\mu})^{n_1} \prod_{j=1}^{n_1} \frac{e^{-i\widehat{\mathbf{k}}\cdot \mathbf{l}_j(t_x-t_y)}-1}{-i\widehat{\mathbf{k}}\cdot \mathbf{l}_j}
\end{split}
\eeq
after summing over all possible permutations of \(n_1\) background field insertions. Similarly,
the lower quark rail with momentum \(\mathbf{k}\) becomes
\beq
\label{Eq:lower_rail}
\begin{split}
S_{\adv}^{n_2}(t_y,t_x;\mathbf{p}+\mathbf{k} + \sum \mathbf{l}_i; \left\{\tilde{\mathbf{l}}_1,\dots \right\}) \\
= S_{\ret}^{(0)}(t_y,t_x;\mathbf{p}+\mathbf{k})  e^{-i\sum_i \widehat{\mathbf{k}}\cdot\mathbf{l}_i(t_y-t_x)}\\
\times (igA_{\mu}\widehat{K}^{\mu})^{n_2} \prod_{j=1}^{n_2} \frac{e^{-i\widehat{\mathbf{k}}\cdot \tilde{\mathbf{l}}_j(t_y-t_x)}-1}{-i\widehat{\mathbf{k}}\cdot \tilde{\mathbf{l}}_j}.
\end{split}
\eeq
with \(n_2\) instability insertions.
The extra factor of \(e^{-i\sum_i \widehat{\mathbf{k}}\cdot\mathbf{l}_i(t_y-t_x)}\) arises because he momentum flow into the advanced propagator is \(\mathbf{p}+\mathbf{k} + \sum \mathbf{l}_i\)  where \(\mathbf{l}_i\) come from the retarded propagator.

When pairing up background insertions we must integrate over the momenta in \(rr\) propagators, \(\mathbf{l}\) and \(\widetilde{\mathbf{l}}\). Pairing up two background fields insertions on the upper quark rail gives a factor
\beq
\label{Eq:D_factor}
\begin{split}
D:&= -g^2 \widehat{K}_{\mu} \widehat{K}_{\nu}\int \frac{d^3l}{(2\pi)^3}\;G_{rr}^{\mu\nu}(T;\mathbf{l}) \;\\
&\qquad\qquad\times \frac{e^{-i\widehat{\mathbf{k}}\cdot \mathbf{l}(t_x-t_y)}-1}{-i\widehat{\mathbf{k}}\cdot \mathbf{l}} \frac{e^{i\widehat{\mathbf{k}}\cdot \mathbf{l}(t_x-t_y)}-1}{i\widehat{\mathbf{k}}\cdot \mathbf{l}} \\
& \approx -g^2 \widehat{K}_{\mu} \widehat{K}_{\nu}\int \frac{d^3l}{(2\pi)^3}\;G_{rr}^{\mu\nu}(T;\mathbf{l}) \left[ (\Delta t)^2 -\frac{1}{12} l^2 (\Delta t)^4\right]
\end{split}
\eeq
where we used that the momentum flow  is \(\mathbf{l}\) in one insertion, and \(-\mathbf{l}\) in the other insertion. The first term, \((\Delta t)^2\), describes a phase shift and the second term describes how the dispersion relation changes because of fluctuating background fields.
Pairing up two background field insertions on the lower quark rail gives the same factor \(D\).
Finally, pairing up an insertion from the upper rail and an insertion from the lower rail gives 
\beq
\begin{split}
-g^2 \widehat{K}_{\mu} \widehat{K}_{\nu} \int \frac{d^3l}{(2\pi)^3} \;&G_{rr}^{\mu\nu}(T;\mathbf{l}) \; \frac{e^{-i\widehat{\mathbf{k}}\cdot \mathbf{l}(t_x-t_y)}-1}{-i\widehat{\mathbf{k}}\cdot \mathbf{l}}\\
&\times e^{-i\widehat{\mathbf{k}}\cdot \mathbf{l}(t_y-t_x)}\frac{e^{i\widehat{\mathbf{k}}\cdot \mathbf{l}(t_y-t_x)}-1}{i\widehat{\mathbf{k}}\cdot \mathbf{l}} %\\
%g^2 \widehat{K}_{\mu} \widehat{K}_{\nu} \int \frac{d^3l}{(2\pi)^3} \frac{e^{-i\widehat{\mathbf{k}}\cdot \mathbf{l}(t_x-t_y)}-1}{-i\widehat{\mathbf{k}}\cdot \mathbf{l}} \frac{e^{i\widehat{\mathbf{k}}\cdot \mathbf{l}(t_x-t_y)}-1}{i\widehat{\mathbf{k}}\cdot \mathbf{l}} = -D
\end{split}
\eeq
which has the value \(-D\).

We must now sum over all possible ways of attaching background field insertions to the two quark rails. 
% We can now evalute photon emission in the background of long wavelength Abelian fields by summing over all possible ways of attaching the background field insertions.
A typical diagram can be seen in Fig. \ref{Fig:crossed_diagr}. Fortunately, we have already summed over all ways of ordering field insertions on each quark rail in Eqs. \eqref{Eq:upper_rail} and \eqref{Eq:lower_rail}.  Thus we only need to sum over the number of insertions on each rail and the different ways of joining them in \(rr\) propagators.
% we are implicitly summing over all possible diagrams when summing over different values of \(m_1, m_2, m_3\).
% Of the \(n\) pairs there are \(m_1\) where both ends are on the quark propagator with momentum \(\mathbf{p}+\mathbf{k}\), \(m_2\) pairs with both ends on the other quark propagator with momentum \(\mathbf{p}\), and \(m_3\) pairs that join the two propagator.
Assuming that there are \(m_1\) \(rr\) propagators where both ends are on the upper quark rail, \(m_2\) propagators with both ends on the lower quark rail and \(m_3\) pairs that join the two propagator, the time dependence becomes
\beq
\label{Eq:combinatorics}
\begin{split}
&\theta(t_x-t_y)e^{-i(\left|\mathbf{p}+\mathbf{k} \right|-p)(t_x-t_y)} \\ &\times\sum_{n=0}^{\infty} \sum_{\substack{ m_1,m_2,m_3\geq 0\\m_1+m_2+m_3 = n }}  \frac{1}{m_1! m_2! m_3!} \frac{1}{2^{m_1}2^{m_2}} D^{m_1} D^{m_2} (-D)^{m_3} 
\end{split}
\eeq
There is a total of \(2n\) background field insertions.  The combinatorial factors account for that the diagram remains the same after interchanging different propagators between the same rails or interchanging the ends of a propagator. (We do not divide by \(2^{m_3}\) since we have not permuted instability insertions between the rails.)
% We have already summed over all permutations on the upper quark rail and the lower quark rail seperately.
% Thus, we must divide by \(2^{m_1} 2^{m_2}\) to account for the fact that interchanging the two vertices in a pair does not change the diagram. (We don't divide by \(2^{m_3}\) since we have not permuted instability insertions between the rails.) Furthermore, we divide by \(m_1 ! m_2 ! m_3 !\) because interchanging the different pairs does not change the diagram. 
The combinatorial sum in Eq. \eqref{Eq:combinatorics} gives 
\beq
\label{Eq:inst_no_effect}
\begin{split}
% B_0 \longrightarrow &B_0 \sum_{n=0}^{\infty} \sum_{\substack{ m_1,m_2,m_3\geq 0\\m_1+m_2+m_3 = n }} \frac{1}{2^n n!} \frac{n!}{m_1! m_2! m_3!} \\
% &\qquad \qquad \qquad \times 2^{m_3} D^{m_1} D^{m_2} (-D)^{m_3} \\
\sum_{n=0}^{\infty} \frac{1}{2^n n!} \left(D+D-2D \right)^n  = 1
\end{split}
\eeq
Thus the total contribution of the background field cancels out in the Abelian case. The same cancellation takes place in the other channels, namely for an antiquark emitting a photon and in quark-antiquark pair annihilation.

%ADD INTERPRETATION

\begin{figure}
    \centering
    \includegraphics[width=0.4\textwidth]{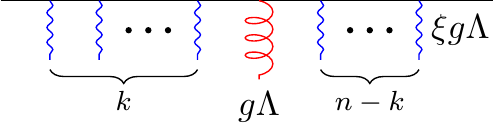}
    \caption{A retarded propagator with one HTL medium kick and \(n\) background field insertions.}   
	\label{Fig:one_HTL}     
\end{figure}

The effect of a long-wavelength background field on photon emission is vanishing to our order of approximation. This is true in the absence of a medium but it turns out to be equally true when there is a medium kicking the quarks back and forth in the transverse plane. To show this, we need retarded propagators including both medium kicks and the effect of a long-wavelength background field.
% So far we have only considered the effect of long wavelength background fields on photon emission when there is no medium. We will now see the effect of background fields on medium-induced splitting including the LPM effect. We show that the effect of the background fields can be factorized and is vanishing in the Abelian case just like above.  We must first rederive quark propagators when there is both a medium and background fields.  
With a medium kick at time \(t\) and
% \(n\) background field insertions either before or after the kick. We sum over all ways of attaching these insertions to the propagator. We first consider the case of 
\(k\) background field insertions before the kick and \(n-k\) insertions after the kick, see Fig. \ref{Fig:one_HTL}, the quark propagator becomes
%\beq
%\label{Eq:rr_bare_2}
%\begin{split}
%&S^{(k)}_{\ret}(t_x,t;\mathbf{p}; \{\mathbf{l}_1,\dots,\mathbf{l}_k\}) \;\mathcal{I}(t;\mathbf{q})\; S^{(n-k)}_{\ret}(t,t_y;\mathbf{p}+\mathbf{q}+\mathbf{l}_1+\dots +  \mathbf{l}_k; \{\mathbf{l}_{k+1},\dots,\mathbf{l}_{n}\}) \\
%&= S^{(0)}_{\ret}(t_x,t;\mathbf{p}) \;\mathcal{I}(t;\mathbf{q})\; S^{(0)}_{\ret}(t,t_y;\mathbf{p}+\mathbf{q}) \\
%&\qquad\qquad\times (ig A_{\mu} \widehat{P}^{\mu})^{n} \prod_{i=1}^k \frac{e^{-i\,\widehat{\mathbf{p}}\cdot\mathbf{l}_i(t_x-t)}-1}{-i\,\widehat{\mathbf{p}}\cdot\mathbf{l}_i} e^{-i\,\widehat{\mathbf{p}}\cdot\mathbf{l}_i(t-t_y)} \prod_{j=k+1}^{n}\; \frac{e^{-i\,\widehat{\mathbf{p}}\cdot\mathbf{l}_j(t-t_y)}-1}{-i\,\widehat{\mathbf{p}}\cdot\mathbf{l}_j}
%\end{split}
%\eeq
\beq
\label{Eq:rr_bare_2}
\begin{split}
&S^{(k)}_{\ret}(t_x,t;\mathbf{p}; \{\mathbf{l}_1,\dots,\mathbf{l}_k\}) \;\mathcal{I}(t;\mathbf{q})\;\\
&\qquad\qquad  \times S^{(n-k)}_{\ret}(t,t_y;\mathbf{p}+\mathbf{q}+\mathbf{l}_1+\dots +  \mathbf{l}_k; \{\mathbf{l}_{k+1},\dots,\mathbf{l}_{n}\}) \\
&= S^{(0)}_{\ret}(t_x,t;\mathbf{p}) \;\mathcal{I}(t;\mathbf{q})\; S^{(0)}_{\ret}(t,t_y;\mathbf{p}+\mathbf{q})\; \\
&\qquad\qquad\times(ig A_{\mu}\widehat{P}^{\mu})^{n} \; \prod_{i=1}^k X(\mathbf{l}_i) \prod_{j=k+1}^{n}\; Y(\mathbf{l}_i)
\end{split}
\eeq
where \(\mathcal{I}\) is the vertex factor for the medium kick and \(\mathbf{q}\) is the momentum flow in the kick.
% Here we focus on the part with \(\widehat{P}\) and have used Eq. \eqref{Eq:instab_perm}. 
% The corresponding diagram can be seen in Fig. \ref{Fig:one_HTL}.
 We have defined 
\beq
% \begin{split}
X(\mathbf{l}_i) = \frac{e^{-i\,\widehat{\mathbf{p}}\cdot\mathbf{l}_i(t_x-t)}-1}{-i\,\widehat{\mathbf{p}}\cdot\mathbf{l}_i} e^{-i\,\widehat{\mathbf{p}}\cdot\mathbf{l}_i(t-t_y)} \\
% &= \frac{e^{-i\,\widehat{\mathbf{p}}\cdot\mathbf{l}_j(t_x-t_y)}-e^{-i\,\widehat{\mathbf{p}}\cdot\mathbf{l}_j(t-t_y)}}{-i\,\widehat{\mathbf{p}}\cdot\mathbf{l}_j}
% \end{split}
\eeq
for background field insertions left of the kick and
\beq
Y(\mathbf{l}_j) = \frac{e^{-i\,\widehat{\mathbf{p}}\cdot\mathbf{l}_j(t-t_y)}-1}{-i\,\widehat{\mathbf{p}}\cdot\mathbf{l}_j}
\eeq
for insertions right of the kick.
We have furthermore used that \(\widehat{\mathbf{p}\!+\!\mathbf{q}} \approx \widehat{\mathbf{p}}\) to avoid corrections of order \(g l (\Delta t)^2\) to Eq. \eqref{Eq:rr_bare_2}.

We now sum over all ways of attaching the \(n\) background field insertions, either before or after the medium kick.
% In Eq. \eqref{Eq:rr_bare_2} we have already summed over permutations of the background field insertions on the left and the right side of the kick separately. Allowing any insertion to be on either side of the medium kick while holding the total number of insertions fixed at \(n\) amounts to evaluating
This gives
\beq
\begin{split}
&S^{(0)}_{\ret}(t_x,t;\mathbf{p}) \;\mathcal{I}(t)\; S^{(0)}_{\ret}(t,t_y;\mathbf{p} + \mathbf{q}) \;\\
&\qquad\qquad\times(ig A_{\mu} \widehat{P}^{\mu})^{n} \prod_{i=1}^{n} \left[ X(\mathbf{l}_i) + Y(\mathbf{l}_j) \right] \\
& = S^{(0)}_{\ret}(t_x,t;\mathbf{p}) \;\mathcal{I}(t)\; S^{(0)}_{\ret}(t,t_y;\mathbf{p} + \mathbf{q}) \;\\
&\qquad\qquad\times(ig A_{\mu} \widehat{P}^{\mu})^{n} \prod_{i=1}^{n} \frac{e^{-i\,\widehat{\mathbf{p}}\cdot\mathbf{l}_i(t_x-t_y)}-1}{-i\,\widehat{\mathbf{p}}\cdot\mathbf{l}_i}
\end{split}
\eeq
The effect of the instabilities is in a factor which does not depend on the time of the medium kick. 
% has been completely factored out into  
% \beq
% (ig A_{\mu} \widehat{P}^{\mu})^{n} \prod_{i=1}^{n} \frac{e^{-i\,\widehat{\mathbf{p}}\cdot\mathbf{l}_i(t_x-t_y)}-1}{-i\,\widehat{\mathbf{p}}\cdot\mathbf{l}_i}
% \eeq
% which does not depend on the time of the medium kick. 
This argument can clearly be extended for any number of medium kicks. Thus, the effect of medium kicks and \(n\) background field insertions factorizes and the dependence on the background field strength is exactly the same as in the case without a medium.
% we can factor out the effect of the background field into 
% \beq
% (ig A_{\mu} \widehat{K}^{\mu})^{n} \prod_{i=1}^{n} \frac{e^{-i\,\widehat{\mathbf{k}}\cdot\mathbf{l}_i(t_x-t_y)}-1}{-i\,\widehat{\mathbf{k}}\cdot\mathbf{l}_i}
% \eeq
% for the \(\widehat{K}\) part and  
% \beq
% (-1)^n(ig A_{\mu} \widetilde{K}^{\mu})^{n} \prod_{i=1}^{n} \frac{e^{i\,\widehat{\mathbf{k}}\cdot\mathbf{l}_i(t_x-t_y)}-1}{i\,\widehat{\mathbf{k}}\cdot\mathbf{l}_i}
% \eeq
% for the \(\widetilde{K}\) part. This applies equally for advanced propagators. For the \(rr\) propagator we can use Eq. \eqref{Eq:rr_F} to factor out the instability insertions (see Appendix \ref{sec:Appendix_distr} for further details). Since we can factor the background field effect out of all propagators we can factor them out of whole diagrams.
The same argument that lead to Eq. \eqref{Eq:inst_no_effect} then shows that the background field does not affect photon emission at leading and next-to-leading order in \(l\Delta t\) and the rate is given by Eq. \eqref{eq:photonrate2}.

\section{Conclusions}

Non-equilibrium QCD plasmas at weak coupling contain instabilities which lead to exponential growth in soft gluon density with time. This makes the plasma inherently time-dependent. We thus argue that quantum field theory calculations that assume a static non-equilibrium plasma do not work. In particular, the rate of medium-induced jet splitting or photon production cannot be evaluated in a static plasma out of equilibrium.

To resolve this affair, we have calculated the time dependence of soft gluon correlators in a simple setup with a slightly anisotropic initial momentum distribution of quasiparticles. Using tools of non-equilibrium quantum field theory, we have derived the retarded correlator, Eq. \eqref{Eq:ret_final}, and \(rr\) correlator, Eq. \eqref{Eq:rr_final}, at early times. As expected, the \(rr\) correlator shows exponential growth in the soft gluon density with time because of instabilities. Using this correct correlator leads to finite and well-behaved rates for medium-induced photon production and jet splitting. The rates depend on the density of gluons in a fluctuating cloud that is sourced at each instant by quasiparticles, as well as the density in instability modes which changes with time. Including the effect of the time-dependent instability modes is difficult in general, but we show that in an Abelian plasma the instabilities' effect on photon production vanishes up to next-to-leading order in \(l \Delta t \ll 1\) where \(l\) is the energy scale of the instabilities and \(\Delta t\) is the time needed to emit a photon.

In phenomenological applications, numerical work using classical-statistical field theory suggests that instability modes are no longer highly occupied once the hydrodynamical stage is reached \cite{Berges:2013eia,Berges:2013fga}. Thus we suggest that one can simply subtract the instability contribution, when calculating photon production or jet-medium interaction  in the non-equilibrium plasma produced in heavy-ion collisions. This gives well-behaved rates that only depend on the instantaneous properties of the medium, 
% Even though the effect of \(\xi g\Lambda\) excitations vanishes there are still important non-equilibrium effects coming from the \(g\Lambda\) kicks. These kicks describe how the emitting particle is deflected in the fields radiated by non-equilibrium quarks and gluons. We have derived a general expression describing this physics, 
see Eq. \eqref{eq:photonrate2}. They contain both non-equilibrium momentum distributions and the non-equilibrium, fluctuating soft gluon cloud. The rate equations can then be solved numerically whenever a momentum distributions of quarks and gluons is specified, see \cite{Hauksson:2018jog,Hauksson2020}. Thus it becomes possible to calculate the effect of shear viscous flow on jet evolution in the plasma, as well as photon production, through all leading-order channels. Combined with a hydrodynamical model of the QGP fluid, this might make it possible to constrain the viscosity of the QGP using jet physics and photons.

% Combining it with a hydrodynamical simulation which specifies the momentum distribution in each fluid cell allows for non-equilibrium jet physics. {\bf This makes possible for the first time to see the effect of the shear viscosity of QGP on jets} and possibly even to constrain the viscosity using jet physics.

\acknowledgements
This work was supported in part by the Natural Sciences and Engineering Research Council of Canada. S. H. gratefully acknowledges a scholarship from the Fonds de recherche du Qu\'ebec - Nature et technologies.

\bibliography{references}

%\begin{thebibliography}{999}
%
%
%\end{thebibliography}

% \newpage

\appendix

% \section{Correlators on a Keldysh-Schwinger contour}
% \label{sec:Appendix_KS}

% ADD
% Include how to have soft gluons in initial state.

\section{Evaluation of correlators}
\label{sec:Appendix_rr}

We begin by deriving the retarded propagator in Eq. \eqref{Eq:ret_final} in full detail. We start from Eq. \eqref{Eq:ret_eval} and evaluate it term by term by inserting principal values as in Eq. \eqref{Eq:princ_value}. The first term becomes 
\beq
\begin{split}
&-\int \frac{dk_1}{2\pi} \int \frac{dk_2}{2\pi} \int_{\alpha} \frac{dk_3}{2\pi} \; G_{\ret}^0(k_1)\, \Pi_{\ret}(k_2)\, G_{\ret}(k_3)\,e^{-ik_1 (x-y)}\\
& \times \frac{1}{8} \Bigg[\left( \frac{i}{k_1-k_2+i\epsilon_1-i\epsilon_2} + \frac{i}{k_1-k_2+i\epsilon_1+i\epsilon_2}\right)\\
&\quad\quad\quad\times\left( \frac{i}{k_1-k_3+i\epsilon_1-i\epsilon_3} + \frac{i}{k_1-k_3+i\epsilon_1+i\epsilon_3}\right) \\
&\;\;\;\;\;+\left(\frac{i}{k_1-k_2-i\epsilon_1-i\epsilon_2} + \frac{i}{k_1-k_2-i\epsilon_1+i\epsilon_2} \right)\\
&\quad\quad\quad\times\left( \frac{i}{k_1-k_3-i\epsilon_1-i\epsilon_3}+\frac{i}{k_1-k_3-i\epsilon_1+i\epsilon_3}\right)  \Bigg]
\end{split}
\eeq
We can continue the \(k_3\) integral by adding a semicircle in the upper half plane with a large radius, see Fig. \ref{Fig:countour}b. Using the residue theorem we avoid all poles of \(G_{\ret}\) so only the poles in the square bracket contribute. We similarly continue the \(k_2\) to the upper half plane. This gives
\beq
\label{Eq:ret_res_a}
\begin{split}
&\int_{\mathbb{R}} \frac{dk_1}{2\pi} \; G_{\ret}^0(k_1)\, \Pi_{\ret}(k_1)\, G_{\ret}(k_1)\,e^{-ik_1 (x-y)}\\
& \times\frac{1}{8} \left[\theta(\epsilon_2-\epsilon_1) \theta(\epsilon_3-\epsilon_1)+ \big(1+\theta(\epsilon_1-\epsilon_2) \big) \big(1+\theta(\epsilon_1-\epsilon_3) \big) \right].
\end{split}
\eeq
where the integration contour is simply the real line.
In a similar fashion, the second term in Eq. \eqref{Eq:ret_eval} becomes 
\beq
\label{Eq:ret_res_b}
\begin{split}
&\int_{\mathbb{R}} \frac{dk}{2\pi} \; G_{\ret}^0(k)\, \Pi_{\ret}(k)\, G_{\ret}(k)\,e^{-ik (x-y)}\\
& \times\frac{1}{8} \left[\theta(\epsilon_1-\epsilon_2) \theta(\epsilon_3-\epsilon_2)  + \big(1+\theta(\epsilon_2-\epsilon_1) \big) \big(1+\theta(\epsilon_2-\epsilon_3) \big)\right].
\end{split}
\eeq 
The third term in Eq. \eqref{Eq:ret_eval} is slightly trickier to evalute. We write the \(k_3\) integration contour as \(\int_{\alpha} = \int_{\mathbb{R}} + \sum_i \int_{\gamma_i}\) where \(\gamma_i\) go around the instability poles in the upper half complex plane, see Fig. \ref{Fig:countour}c. The part with the real line integration gives   
\beq
\label{Eq:ret_res_c}
\begin{split}
&\int_{\mathbb{R}} \frac{dk_3}{2\pi} \; G_{\ret}^0(k_3)\, \Pi_{\ret}(k_3)\, G_{\ret}(k_3)\,e^{-ik_3 (x-y)}\\
& \times\frac{1}{8} \left[ \left(1+\theta(\epsilon_3-\epsilon_1) \right) \left(1+\theta(\epsilon_3-\epsilon_2) \right) + \theta(\epsilon_1-\epsilon_3) \theta(\epsilon_2-\epsilon_3)\right]
\end{split}
\eeq
after doing the \(k_1\) and \(k_2\) integrals. The part with integration over the \(\gamma_i\) contour can be done explicitly giving 
\beq
\label{Eq:ret_res_d}
\sum_i \int_{\gamma_i} \frac{dk}{2\pi}\; G_{\ret}^0(k)\, \Pi_{\ret}(k)\, G_{\ret}(k) e^{-ik(x-y)}.
\eeq

We are finally in a position to find the retarded propagator. Adding up the contributions of Eqs. \eqref{Eq:ret_res_a}, \eqref{Eq:ret_res_b}, \eqref{Eq:ret_res_c} and \eqref{Eq:ret_res_d} and using identities of \(\theta\)-functions we get
\beq
\int_{\alpha} \frac{dk}{2\pi}\; G_{\ret}^0(k)\, \Pi_{\ret}(k)\, G_{\ret}(k) e^{-ik(x-y)}
\eeq
which leads directly to Eq. \eqref{Eq:ret_from_appendix} and thus to  Eq. \eqref{Eq:ret_final} as we wanted to show.

%%%%%%%%%%%%%%%%%%%%%%%%%%%%%%%%%%%%%%%%
%%%%%%%%%%%%%%%%%%%%%%%%%%%%%%%%%%%%%%%%%%

We next evaluate the \(rr\) correlator. Specifically, we will show how Eq. \eqref{Eq:rr_result} follows from Eq. \eqref{Eq:rr_prop} using the approximations described in Chapter \ref{sec:correlators}. Just like for the evaluation of \(G_{\ret}\) there are no poles when \(k_1 = k_2\) or \(k_2 = k_3\) which allows us to insert principal values. This gives that 
\begin{widetext}
\beq
\label{Eq:app_rr_full}
\begin{split}
G_{rr}&(x,y) = \int_{\alpha} \frac{dk_1}{2\pi} \int \frac{dk_2}{2\pi} \int_{\widetilde{\alpha}} \frac{dk_3}{2\pi} \; \left[e^{-ik_2(x-y)} - e^{-ik_1x}e^{ik_2 y}-e^{-ik_2 x}e^{ik_3 y} + e^{-ik_1 x}e^{ik_3 y}\right]  \\
& \times\frac{1}{8} \Bigg[ \left(\frac{1}{k_2-k_1 + i\epsilon_2 - i \epsilon_1} + \frac{1}{k_2-k_1 + i\epsilon_2 + i \epsilon_1} \right) \left( \frac{1}{k_2-k_3 + i\epsilon_2 - i \epsilon_3} +  \frac{1}{k_2-k_3 + i\epsilon_2 + i \epsilon_3}\right) \\
& +\left(\frac{1}{k_2-k_1 - i\epsilon_2 - i \epsilon_1} + \frac{1}{k_2-k_1 - i\epsilon_2 + i \epsilon_1} \right) \left( \frac{1}{k_2-k_3 - i\epsilon_2 - i \epsilon_3} +  \frac{1}{k_2-k_3 - i\epsilon_2 + i \epsilon_3}\right)  \Bigg]\\
&\qquad\times \left(\widehat{G}_{\ret}(k_1) + \sum_{i} \frac{A_i}{k_1 - a_i} \right) \, \Pi_{aa}(k_2)\, \left( \widehat{G}_{\adv}(k_3) + \sum_{j} \frac{A^*_j}{k_3 - a_j^*}\right).
\end{split}
\eeq
\end{widetext}
where we have substituted the scale sepaparation of Eqs. \eqref{Eq:ret_scale_sep} and \eqref{Eq:adv_scale_sep}.

The evaluation of Eq. \eqref{Eq:app_rr_full} depends on the scale one is working at. We begin by evaluating terms at the scale \(g\Lambda\), i.e. terms with \(\widehat{G}_{\ret}\) and \(\widehat{G}_{\adv}\).  We do this one exponential at a time. The first exponential term (i.e. all terms with \(e^{-ik_2(x-y)}\)) can be evaluated exactly by continuing the \(k_1\) integral to the upper half complex plane, the \(k_3\) to the lower half complex plane and applying the residue theorem. Then all poles of \(\widehat{G}_{\ret}\) and \(\widehat{G}_{\adv}\) are avoided and one gets
\beq
\label{Eq:app_rr_reg_a}
\begin{split}
&\int \frac{dk_2}{2\pi}\;  \widehat{G}_{\ret}(k_2)\, \Pi_{aa}(k_2)\, \widehat{G}_{\adv}(k_3)\;\; e^{-ik_2(x-y)}\\
&\times\frac{1}{8} \left[\theta(\epsilon_3 - \epsilon_2) \big( 1+\theta(\epsilon_2-\epsilon_1)\big)  + \theta(\epsilon_1-\epsilon_2) \big(1+\theta(\epsilon_2-\epsilon_3) \big) \right].
\end{split} 
\eeq
In the second exponential term (i.e. all terms with \(-e^{-ik_1x}e^{ik_2 y}\)) in Eq. \eqref{Eq:app_rr_full} we continue the \(k_3\) integral to the lower half plane giving
\beq
\begin{split}
&-i \int_{\alpha} \frac{dk_1}{2\pi} \int \frac{dk_2}{2\pi}\;\widehat{G}_{\ret}(k_1)\, \Pi_{aa}(k_2)\, \widehat{G}_{\adv}(k_2) \;\;e^{-ik_1 x} \,e^{ik_2 y} \\
&\times\frac{1}{8} \Bigg[ \theta(\epsilon_3 - \epsilon_2) \left(\frac{1}{k_2-k_1+i\epsilon_2-i\epsilon_1} + \frac{1}{k_2-k_1+i\epsilon_2+i\epsilon_1} \right)  \\
&+\big(1+\theta(\epsilon_2 - \epsilon_3)\big) \\
&\times\left(\frac{1}{k_2-k_1-i\epsilon_2-i\epsilon_1} + \frac{1}{k_2-k_1-i\epsilon_2+i\epsilon_1} \right)\Bigg].
\end{split}
\eeq
exactly. In order to evalute the \(k_1\) integral we need to use our approximations.  Because of the exponential we must continue the contour to the lower half complex plane. Applying the residue theorem we get a contribution from all poles and branch cuts of \(\widehat{G}_{\ret}\) but they all contain a factor \(e^{-ibx}\) with \(b\sim g\Lambda\) and can thus be dropped according to our approximations. Thus only poles with \(k_2 = k_1\) contribute, giving
\beq
\label{Eq:app_rr_reg_b}
\begin{split}
&\int \frac{dk_2}{2\pi}\; \widehat{G}_{\ret}(k_2)\, \Pi_{aa}(k_2)\, \widehat{G}_{\adv}(k_2) \;\;e^{-ik_2 (x-y)}\\
&\times \frac{1}{8} \Big[ \theta(\epsilon_1-\epsilon_2)\theta(\epsilon_3-\epsilon_2) \\
&\qquad \qquad + \big(1+\theta(\epsilon_2-\epsilon_3)\big)\big(1+\theta(\epsilon_2-\epsilon_1)\big)\Big]. 
\end{split}
\eeq
In the same way, the third exponential term in Eq. \eqref{Eq:app_rr_full} is 
\beq
\label{Eq:app_rr_reg_c}
\begin{split}
&\int \frac{dk_2}{2\pi}\; \widehat{G}_{\ret}(k_2)\, \Pi_{aa}(k_2)\, \widehat{G}_{\adv}(k_2) \;\;e^{-ik_2 (x-y)} \\
&\times \frac{1}{8} \Big[ \theta(\epsilon_1-\epsilon_2)\theta(\epsilon_3-\epsilon_2)\\
&\qquad \qquad +\big(1+\theta(\epsilon_2-\epsilon_1)\big)\big(1+\theta(\epsilon_2-\epsilon_3)\big)\Big]
\end{split}
\eeq
and the fourth exponential is
\beq
\label{Eq:app_rr_reg_d}
\begin{split}
&\int \frac{dk_2}{2\pi}\; \widehat{G}_{\ret}(k_2)\, \Pi_{aa}(k_2)\, \widehat{G}_{\adv}(k_2) \;\;e^{-ik_2 (x-y)}\\
&\times \frac{1}{8} \Big[ \theta(\epsilon_1-\epsilon_2)\big(1+\theta(\epsilon_2-\epsilon_3)\big) \\
&\qquad\qquad+ \theta(\epsilon_3-\epsilon_2)\big(1+\theta(\epsilon_2-\epsilon_1)\big)\Big]. 
\end{split}
\eeq
Adding up the different terms in Eqs. \eqref{Eq:app_rr_reg_a}, \eqref{Eq:app_rr_reg_b}, \eqref{Eq:app_rr_reg_c}, \eqref{Eq:app_rr_reg_d} and using identities for \(\theta\)-functions, we get that the contribution to the \(rr\) propagator at the scale \(g\Lambda\) is 
\beq
\approx \int \frac{dk}{2\pi}\; \widehat{G}_{\ret}(k) \,\Pi_{aa}(k)\, \widehat{G}_{\adv}(k) \;\;e^{-ik(x-y)}.
\eeq

We next turn to evaluating terms in Eq. \eqref{Eq:app_rr_full} at the scale \(\xi g\Lambda\), i.e. the contribution of instability poles in the retarded and advanced propagators. As before the contribution of the first exponential is 
\beq
\label{Eq:app_rr_inst_a}
\begin{split}
\sum_{i,j}&\int \frac{dk_2}{2\pi}\;  \frac{A_i}{k_2-a_i}\, \Pi_{aa}(k_2)\, \frac{A_j^*}{k_2-a_j^*}\;\; e^{-ik_2(x-y)}\\
&\times\frac{1}{8} \Big[\theta(\epsilon_3 - \epsilon_2) \big( 1+\theta(\epsilon_2-\epsilon_1)\big)  \\
&\qquad\qquad+ \theta(\epsilon_1-\epsilon_2) \big(1+\theta(\epsilon_2-\epsilon_3) \big) \Big].
\end{split} 
\eeq
In the second exponential in Eq. \eqref{Eq:app_rr_full} we continue the \(k_3\) integral to the lower half plane to get 
\begin{widetext}
\beq
\begin{split}
-i \sum_{i,j} &\int_{\alpha} \frac{dk_1}{2\pi} \int \frac{dk_2}{2\pi}\; \frac{A_i}{k_1-a_i}\, \Pi_{aa}(k_2)\, \frac{A_j^*}{k_2-a_j^*}\;\;e^{-ik_1 x} \,e^{ik_2 y} \\
&\times\frac{1}{8} \Bigg[ \theta(\epsilon_3 - \epsilon_2) \left(\frac{1}{k_2-k_1+i\epsilon_2-i\epsilon_1} + \frac{1}{k_2-k_1+i\epsilon_2+i\epsilon_1} \right)  \\
&\qquad\;\;\big(1+\theta(\epsilon_2 - \epsilon_3)\big) \left(\frac{1}{k_2-k_1-i\epsilon_2-i\epsilon_1} + \frac{1}{k_2-k_1-i\epsilon_2+i\epsilon_1} \right)\Bigg].
\end{split}
\eeq
\end{widetext}
Now when we continue the \(k_1\) integral to the lower half plane we get a contribution from \(k_1 = k_2\) as well as a contribution from \(k_1 = a_i\) leading to
\beq
\label{Eq:app_rr_inst_b}
\begin{split}
&\sum_{i,j} \int \frac{dk_2}{2\pi}\; \frac{A_i}{k_2-a_i}\, \Pi_{aa}(k_2)\, \frac{A_j^*}{k_2-a_j^*}\;e^{-ik_2 (x-y)} \\
&\times \frac{1}{8} \left[ \theta(\epsilon_1-\epsilon_2)\theta(\epsilon_3-\epsilon_2) + (1+\theta(\epsilon_2-\epsilon_3))(1+\theta(\epsilon_2-\epsilon_1))\right] \\
&- \frac{1}{2} \sum_{i,j} \int \frac{dk_2}{2\pi} \;\frac{A_i}{k_2-a_i}\, \Pi_{aa}(k_2)\, \frac{A_j^*}{k_2-a_j^*} \;\;e^{-ia_i x}\,e^{ik_2 y}
\end{split}
\eeq
Similarly, the third exponential in Eq. \eqref{Eq:app_rr_full} is
\beq
\label{Eq:app_rr_inst_c}
\begin{split}
&\sum_{i,j} \int \frac{dk_2}{2\pi}\; \frac{A_i}{k_2-a_i}\, \Pi_{aa}(k_2)\, \frac{A_j^*}{k_2-a_j^*} \;\;e^{-ik_2 (x-y)} \\
&\times \frac{1}{8} \left[ \theta(\epsilon_1-\epsilon_2)\theta(\epsilon_3-\epsilon_2)+(1+\theta(\epsilon_2-\epsilon_1))(1+\theta(\epsilon_2-\epsilon_3))\right] \\
&- \frac{1}{2} \sum_{i,j} \int \frac{dk_2}{2\pi} \;\frac{A_i}{k_2-a_i}\, \Pi_{aa}(k_2)\, \frac{A_j^*}{k_2-a_j^*} \;\;e^{-ik_2 x}\,e^{ia_j^* y}
\end{split}
\eeq
and the fourth one is
\beq
\label{Eq:app_rr_inst_d}
\begin{split}
\sum_{i,j} &\int \frac{dk_2}{2\pi}\;  \frac{A_i}{k_2-a_i}\, \Pi_{aa}(k_2)\, \frac{A_j^*}{k_2-a_j^*}\;\; e^{-ik_2(x-y)} \\
&\times \frac{1}{8} \Bigg[ \theta(\epsilon_1-\epsilon_2) (1+\theta(\epsilon_2-\epsilon_3))\\ 
& \qquad\qquad+\theta(\epsilon_3-\epsilon_2) (1+\theta(\epsilon_2-\epsilon_1))\Bigg] \\
-\frac{1}{2} \sum_{i,j}&\int \frac{dk_2}{2\pi} \; \frac{A_i}{k_2-a_i}\, \Pi_{aa}(k_2)\, \frac{A_j^*}{k_2-a_j^*}\;\; e^{-ik_2x}\, e^{ia_j^* y}\\
-\frac{1}{2} \sum_{i,j}&\int \frac{dk_2}{2\pi} \; \frac{A_i}{k_2-a_i}\, \Pi_{aa}(k_2)\, \frac{A_j^*}{k_2-a_j^*} \;\;e^{-ia_ix} \,e^{ik_2 y}\\
+ \sum_{i,j}&\int \frac{dk_2}{2\pi} \; \frac{A_i}{k_2-a_i}\, \Pi_{aa}(k_2)\, \frac{A_j^*}{k_2-a_j^*} \;\;e^{-ia_ix} \,e^{ia_j^* y}.
\end{split}
\eeq
Adding up the contributions in Eqs. \eqref{Eq:app_rr_inst_a}, \eqref{Eq:app_rr_inst_b}, \eqref{Eq:app_rr_inst_c}, \eqref{Eq:app_rr_inst_d} then gives that the contribution to \(G_{rr}\) at the scale \(\xi g \Lambda\) is
\beq
\begin{split}
&\sum_{i,j} \int \frac{dk}{2\pi}\; \frac{A_i}{k-a_i} \,\Pi_{aa}(k)\, \frac{A_j^*}{k-a_j^*}  \\
&\times\left(e^{-ikx}-e^{-ia_i x} \right)\left( e^{iky}-e^{ia_j^*y}\right)
\end{split}
\eeq
The calculation for mixed terms with, say, contribution at scale \(g\Lambda\) from the retarded correlator and contribution at scale \(\xi g \Lambda\) from the advanced correlator proceeds analogously.
The final results is precisely Eq. \eqref{Eq:rr_result}.

\section{Evaluation of factors with momentum distributions} %CHANGE
\label{sec:Appendix_distr}

We begin by showing Eq. \eqref{Eq:S_1122}, namely that
\beq
S_{1122} = 2 F(P+K) \left(1-F(P)\right) \mathrm{Re}\; S_{rraa}
\eeq
in the presence of instabilities.  Using that 
\beq
\phi_1 = \phi_r + \frac{1}{2}\phi_a, \qquad \qquad \phi_2 = \phi_r - \frac{1}{2} \phi_a
\eeq
it is easy to see that
\beq
\begin{split}
S_{1122} = S_{rrrr} + \frac{1}{2}\left(S_{arrr} + S_{rarr} - S_{rrar} - S_{rrra} \right) \\
 + \frac{1}{4} \left(S_{aarr} -S_{arar} - S_{arra} -S_{raar} -S_{rara} + S_{rraa} \right) \\
 + \frac{1}{8} \left(-S_{aaar} -S_{aara} + S_{araa} + S_{raaa} \right) \\
 + \frac{1}{16} S_{aaaa}
\end{split}
\eeq
The different four-point functions are defined in Fig. \ref{Fig:4_pt_fun}.
Using that \(aa\) propagators vanish we see that \(S_{naam} = S_{anma} = 0\) for any \(n,m\in \{a,r\}\) so we're then left with 
\beq
\begin{split}
S_{1122} = S_{rrrr} + \frac{1}{2}\left(S_{arrr} + S_{rarr} - S_{rrar} - S_{rrra} \right) \\
 + \frac{1}{4} \left(S_{aarr} -S_{arar}  -S_{rara} + S_{rraa} \right) 
\end{split}
\eeq
Using Eq. \eqref{Eq:rr_F} we furthermore see that \(S_{rnmr} = \left[ \frac{1}{2} - F(P)\right]\left(S_{anmr}-S_{rnma}\right)\) and \(S_{nrrm} = \left[ \frac{1}{2} - F(P+K)\right]\left(S_{nram}-S_{narm}\right)\) so
\begin{widetext}
\beq
\begin{split}
S_{1122} = \left[ \frac{1}{2} - F(P+K)\right] \left[\frac{1}{2} - F(P) \right] \left(-S_{rraa} + S_{arar} + S_{rara} - S_{aarr} \right)  \\
+\left[\frac{1}{2} - F(P+K) \right]\frac{1}{2}\left(S_{arar} - S_{aarr} - S_{rraa} + S_{rara} \right) \\ 
+ \left[\frac{1}{2} - F(P) \right] \frac{1}{2}\left(-S_{rara}  + S_{aarr}+ S_{rraa} - S_{arar}  \right) \\
 + \frac{1}{4} \left(S_{aarr} -S_{arar}  -S_{rara} + S_{rraa} \right) 
\end{split}
\eeq
\end{widetext}
We furthermore have that \(S_{arar} = S_{rara} = 0\) because the two quark propagators give theta functions of the form \(\theta(t_x-t_y)\theta(t_y-t_x) = 0\). We're then left with 
% \beq
% \begin{split}
% S_{1122} = \left[ \frac{1}{2} - F(P+K)\right] \left[\frac{1}{2} - F(P) \right] \left(-S_{rraa}  - S_{aarr} \right)  \\
% +\left[\frac{1}{2} - F(P+K) \right]\frac{1}{2}\left( - S_{aarr} - S_{rraa} \right) \\ 
% + \left[\frac{1}{2} - F(P) \right] \frac{1}{2}\left(   S_{aarr}+ S_{rraa} \right) \\
%  + \frac{1}{4} \left(S_{aarr}  + S_{rraa} \right) 
% \end{split}
% \eeq
% which can also be written as
\beq
\begin{split}
S_{1122} &= F(P+K) \left(1-F(P)\right) \left( S_{rraa}  +S_{aarr}\right) \\
&= 2 F(P+K) \left(1-F(P)\right) \mathrm{Re}\; S_{rraa}
\end{split}
\eeq 

We finally note how the momentum factors work out when there is a medium as well as background fields. Adding \(n\) background field insertions to the bare \(rr\) propagator gives 
\beq
\begin{split}
&S^{(n)}_{rr}(t_x,t_y;\mathbf{p}; \{\mathbf{l}_1,\dots,\mathbf{l}_n\}) \\ &= \left[\frac{1}{2} -   F(\mathbf{k})\right]
\times \Bigg( S^{(n)}_{\ret}(t_x,t_y;\mathbf{p};\{\mathbf{l}_1,\dots,\mathbf{l}_n\}) \\ 
&\qquad\qquad\qquad\qquad- S^{(n)}_{\adv}(t_x,t_y;\mathbf{p}; \{\mathbf{l}_1,\dots,\mathbf{l}_n\}) \Bigg).
\end{split}
\eeq
as in Eq. \eqref{Eq:rr_F}. We can factor out the instability insertions to get 
\beq
\begin{split}
S^{(n)}&_{rr}(t_x,t_y;\mathbf{p}; \{\mathbf{l}_1,\dots,\mathbf{l}_n\}) \\
 &= \left[\frac{1}{2} -   F(\mathbf{k})\right] \left( S^{(0)}_{\ret}(t_x,t_y;\mathbf{p})  - S^{(0)}_{\adv}(t_x,t_y;\mathbf{p}) \right) \\
&\times (igA_{\mu}\widehat{K}^{\mu})^{n} \prod_{j=1}^{n} \frac{e^{-i\widehat{\mathbf{k}}\cdot \mathbf{l}_j(t_x-t_y)}-1}{-i\widehat{\mathbf{k}}\cdot \mathbf{l}_j}
\end{split}
\eeq
or in other words
\beq
\begin{split}
S^{(n)}_{rr}&(t_x,t_y;\mathbf{p}; \{\mathbf{l}_1,\dots,\mathbf{l}_n\}) \\
&= S^{(0)}_{rr}(t_x,t_y;\mathbf{p}) \;(igA_{\mu}\widehat{K}^{\mu})^{n} \prod_{j=1}^{n} \frac{e^{-i\widehat{\mathbf{k}}\cdot \mathbf{l}_j(t_x-t_y)}-1}{-i\widehat{\mathbf{k}}\cdot \mathbf{l}_j}
\end{split}
\eeq
for the \(\widehat{P}\) part and similarly for the \(\widetilde{P}\) part. The same argument as in Chapter \ref{sec:with_medium} then allows us to factor out the effect of background fields for any combination of \(rr\), retarded and advanced propagator and shows that the effect of the background field vanishes. 
% The remaining diagram without any background field insertions can be evaluated with the methods of REF. In particular the overall momentum distribution can be shown to be what we would expect.

\end{document}